\documentclass[twocolumn,showpacs,amsmath,amssymb,floatfix,prb]{revtex4}

\newcommand{\dSum}{\displaystyle\sum}
\usepackage{graphicx}
\usepackage{dcolumn}
\usepackage{bm}

\begin{document}


\title{Structural, mechanical and thermodynamic properties of a coarse-grained DNA model}
\author{Thomas E. Ouldridge$^1$}
\author{Ard A. Louis$^1$}
\author{Jonathan P. K. Doye$^2$}
\affiliation{$^1$Rudolf Peierls Centre for Theoretical Physics, 1 Keble
        Road, Oxford, UK OX1 3NP, UK \\ 
        $^2$Physical \& Theoretical Chemistry Laboratory, Department of Chemistry, University of Oxford, 
  South Parks Road, Oxford, OX1 3QZ, UK}

\date{\today}

\begin{abstract}
We explore in detail the structural, mechanical and thermodynamic properties of a coarse-grained model of DNA similar to  that introduced in Ref.\ \onlinecite{Ouldridge_tweezers_2010}. Effective interactions are used to represent chain connectivity, excluded volume, base stacking and hydrogen bonding, naturally reproducing a range of DNA behaviour. We quantify the relation to experiment of the thermodynamics of  single-stranded stacking, duplex hybridization and hairpin formation, as well as structural properties such as the persistence length of single strands and duplexes, and the torsional and stretching stiffness of double helices. We also explore the model's representation of more complex motifs involving dangling ends, bulged bases and internal loops, and the effect of stacking and fraying on the thermodynamics of the duplex formation transition.
 \end{abstract}

\maketitle

\section{Introduction}
\label{introduction}
A single-stranded molecule of DNA (ssDNA) consists of a chain of alternating sugar and phosphate groups.\cite{Saenger1984} Attached to each sugar is a base, alanine (A), thymine (T), cytosine (C) or guanine (G). Bases are inherently planar, and their tendency to from coplanar stacks and undergo hydrogen-bonding leads to the formation of double-stranded helices (dsDNA). The canonical Watson-Crick base pairs (bp), C-G and A-T, are called complementary base pairs because they form the most stable hydrogen bonds.

The different base identities, along with the rules of complementarity, allow information to be encoded into the single strands.\cite{Watson1953} In nature, this allows both strands of a double helix to carry the genetic information required for life. Recently, this information-carrying property has been harnessed in nanotechnology. A set of single strands can be designed with a pattern of complementarity that specifies a certain 2- or 3-dimensional structure (usually formed from branched double-helices) as the global free-energy minimum of the system. Strands can then be mixed and self-assemble, provided the sequences are well designed. When combined with its structural properties (dsDNA is stiff on the nanoscale, with a persistence length of around 50\,nm or 150 bp,\cite{Hagerman1988} and ssDNA has the flexibility to act as hinges between duplex sections), such selective interactions make DNA an ideal material for nanoscale self-assembly.

The self-assembly of short strands (oligonucleotides) was first demonstrated by the Seeman lab, who created a four-armed junction.\cite{Kallenbach83} Junctions of this type, and more complex motifs,\cite{Fu93, Yan2003} have been used to create lattices\cite{Winfree98, Malo2005} and ribbons.\cite{Yan2003} 3-dimensional structures have also been realized: initially, the Seeman group constructed a cube \cite{Chen91} and a truncated octahedron \cite{Zhang94} in several discrete stages. Polyhedral cages that rapidly form as solutions of oligonucleotides are cooled have since been developed.\cite{Goodman2005, Erben07, Shih04, Anderson08, He2008} These examples illustrate the potential of DNA as a material for controllable nanoscale self-assembly. 

An alternative approach to self-assembly, DNA origami, was recently developed by Rothemund.\cite{Rothemund06} In this case, a long single strand is folded into a desired structure by short ``staple" strands, allowing the assembly of an enormous range of 2-dimensional structures. This approach has  been extended to three dimensions, either by linking together 2-dimensional sheets,\cite{Andersen09} or by using the twist of DNA to form inherently 3-dimensional folded strands.\cite{Douglas09} Additional methods of 3-dimensional self-assembly are possible: self-interactions within a single strand have been used to create a tetrahedron,\cite{Li2009} and other structures have been created from pre-assembled components involving DNA and other organic molecules.\cite{Aldaye2007,Zimmermann08}

DNA nanotechnology is not limited to the self-assembly of static structures, as hybridization can also be used to drive nanodevices.\cite{Bath2007} Such devices typically undergo structural changes due to duplex formation or  toehold-mediated strand displacement (wherein a strand in a partially formed duplex is replaced by a strand which can form a more complete duplex). The earliest designs, such as the ``tweezers" of Yurke {\it et al.},\cite{Yurke2000} required sequential addition of single strands to force a system through a conformational cycle, or along a track.\cite{Sherman2004, Shin2004} The use of enzyme-facilitated hydrolysis,\cite{Bath2009} or fuel in metastable states such as single-stranded hairpins,\cite{Green2008} has allowed the design of autonomous devices. The selectivity of DNA binding has also been used to perform simple logic operations,\cite{Zhang2010} offering the potential for ``intelligent" nanostructures or devices, which respond to certain features of their environment.

As discussed above, much of DNA nanotechnology relies either largely or entirely upon B-DNA duplex hybridization from single strands (although other transitions can be exploited, such as the formation of single-stranded ``i motif" structures\cite{Liedl2005}). Furthermore, some biologically relevant behaviour (such as the opening of transient ``bubbles" (stretches of broken bps)  within helices and the extrusion of cruciform structures in negatively supercoiled DNA\cite{Sinden1994}) relies primarily on the properties of single and double strands, and the competition between the two. 

Information about the intermediate states in assembly processes, which are often difficult to resolve in experiment yet crucial to the processes as a whole, would aid the design of nanostructures and nanotechnology. Computer modelling, provided it can capture the transition between single- and double-stranded DNA, has the potential to offer significant insight into these systems. 

At the most detailed level, atomistic simulations using force fields such as AMBER or CHARM offer an intimate representation of DNA.\cite{Orozco2003}  A large-scale systematic study of the structural properties of short sequences as represented by AMBER has been carried out  by the Ascona B-DNA Consortium.\cite{Lavery2010} Unfortunately, the number of degrees of freedom (including those of the solvating H$_2$O molecules) prohibits the simulation of large molecules for long periods of time. For example, simulations of double helices (on the scale of 10--20 base pairs) have only recently been extended to time scales of $\sim 1\,\mu$s.\cite{Perez2007, Mura2008} The use of enhanced sampling techniques has given atomistic simulations some access to hybridization transitions in the smallest duplexes\cite{Kannan2009} and hairpins,\cite{Sorin2003, Kannan2007} although larger systems remain prohibitively expensive to model. 

At the other end of the spectrum, continuum models of DNA\cite{Marko2005} treat the double helix as a uniform medium. Whilst these approaches can provide important insight into DNA behaviour on long length-scales, they are not constructed to deal with the details of processes involving duplex hybridization or melting.

To gain further insight into hybridization, coarse-grained models, which represent DNA through a reduced set of degrees of freedom with effective interactions, are required. In particular, models whose coarse-grained scale is approximately that of the nucleotide may provide the necessary compromise between resolution and computational speed. 

The simplest available coarse-grained models are statistical, neglecting structural and dynamical detail.
These models use sequence-dependent parameters that describe the free-energy gain per base pair relative to the denatured state, with extra parameters used for initialization of duplex regions and to describe unpaired sections within the a structure. Among the most popular are the Poland-Scheraga\cite{Poland1966} and nearest-neighbour models,\cite{SantaLucia1998, SantaLucia2004} generally used in the context of polynucleotide and oligonucletide melting, respectively. A particularly important variant of the nearest-neighbour model, which has been shown to reproduce experimental melting temperatures of duplexes ranging from 4--16\,bp in length with a standard deviation of $2.3$\,K, was introduced by SantaLucia and Hicks.\cite{SantaLucia1998, SantaLucia2004} In this model, the concentrations of oligonucletides $A$ and $B$, and their duplex $AB$, are given by:
\begin{equation}
\frac{[AB]}{[A][B]} = \exp\Big(-\beta \big(\Delta H_{AB} - T \Delta S_{AB}\big)\Big),
\label{2sm}
\end{equation}
where the constants $\Delta H_{AB}$ and $\Delta S_{AB}$ are computed by summing contributions from each nearest-neighbour set of two base pairs, together with terms for helix initiation and various structural features, all of which are assumed to be temperature independent. Such a description, in which $\Delta H_{AB}$ and $\Delta S_{AB}$ are temperature independent, constitutes a ``two-state" model. 

Alternatives to these purely statistical models have also been proposed. 
Everaers {\it et al.}~\cite{Everaers2007}  have suggested a lattice model of DNA explicitly designed to unify nearest-neighbour and Poland-Scheraga models,with the added advantage that some structural information is also preserved. Peyrard-Bishop-Dauxois models\cite{Dauxois1993} represent base pairs through a continuous 1-dimensional coordinate, allowing dynamical simulations of denaturation bubbles in polynucleotide DNA. None of the models discussed, however, provide a sufficiently sophisticated representation of the three-dimensional structure and dynamics of DNA to allow the detailed study of the transitions involved in nanotechnology.

To study the processes involved in nucleic acid structure formation, a fully 3-dimensional, dynamical, coarse-grained model is required. ``Rigid base-pair" models, in which undeformable base pairs are the fundamental unit, have been used to study perturbations to DNA such as those induced by enzymes.\cite{Becker2009} By definition, such models cannot represent the transition from single strands to duplexes, and hence are inappropriate for the study of assembly processes. Lankas {\it et al.} \cite{Lankas2009} directly compared  rigid base-pair and rigid base models that were parameterized to reproduce positional time-series that were generated from atomistic simulations of B-DNA.  Interestingly, they found that the rigid base models, in which the base pairs are deformable and nucleotides are the essential unit of simulation, generated a more local representation of the interactions than rigid base-pair models did, suggesting that the base-pairs are a more appropriate level of description for structural and mechanical properties of B-DNA.

Several rigid base models, and others in which each nucleotide has stiff internal degrees of freedom, have been proposed in the last decade. These models represent nucleotides by several interaction sites, and can be divided into two kinds. Firstly, some modellers parameterize their effective force fields by direct comparison with either atomistic simulations or data from crystal structures. An alternative is to take a more heuristic approach, designing force fields to provide a reasonable description of a range of large-scale properties (such as melting temperatures of helices) when compared to experiment: these two approaches could be described as ``bottom-up" and ``top-down", respectively.

Bottom-up approaches have been used to study RNA nanostructures,\cite{Paliy2010} the response of DNA minicircles to supercoiling,\cite{Trovato2008, Sayar2009} the behaviour of B-DNA over a range of conditions,\cite{Dans2010} binding of DNA to the nucleosome\cite{Voltz2008} and the properties of the resultant model as a function of parameterization.\cite{Morriss-Andrews2010} Although systematically coarse-graining removes some of the arbitrary choices in designing a minimal model, there are drawbacks. Firstly, the resultant force-field will be biased towards the structures with which it was parameterized: in particular, equilibrium duplex structures are often the primary source of information, and hence single-stranded behaviour is not necessarily well reproduced. Perhaps more significantly, the transition between ssDNA and dsDNA may be poorly represented: indeed, none of the bottom-up approaches described above have been used to investigate melting transitions in a rigorous way, with the focus being largely on structural properties. Secondly, ``representability problems"\cite{Louis2002} mean that careful fitting to distribution functions will not necessarily reproduce thermodynamic properties in a reliable fashion.\cite{Johnson2007} Finally, it is not yet known how accurate atomistic simulations are in reproducing the duplex hybridization transition.

All coarse-grained models represent a compromise, and an appropriate model must be chosen for the investigation at hand. Current examples of bottom-up approaches are well-suited to studying fluctuations in the vicinity of the equilibrium structure in question. By contrast, top-down approaches appear to lend themselves to the study of larger changes, particularly assembly transitions. Top-down approaches have been used to study duplex denaturation,\cite{Drukker2001} hairpin formation,\cite{Sales-Pardo2005, Kenward2009} RNA folding\cite{Ding2008,Pasquali2010} and mechanical unfolding,\cite{Hyeon2005, Hyeon2007} Holliday junction formation,\cite{Ouldridge2009} duplex thermodynamics\cite{Sambriski2008, Sambriski2009} and overstretching.\cite{Niewieczerzal2009}

For this paper we are mainly concerned with developing a model that can treat the formation of complexes involving single strands and  B-DNA, with the particular goal of describing processes that are relevant to the self-assembly of DNA nanostructures and the dynamics of nanodevices,\cite{Ouldridge_tweezers_2010} but also with a view towards biological applications.   We thus require a good representation of the structural, mechanical and thermodynamic properties of both single and double stranded DNA.   

An important property to reproduce is the tendency of consecutive bases tend to form coplanar stacks, with an average separation of about  3.4\,\AA,\cite{Pitchiaya2006}, which is shorter than the equilibrium separation of phosphates (along the backbone) of approximately 6.5\,\AA.\cite{murphy2004}   The difference between the two length-scales helps determine the shape of B-DNA, which forms a double helix to exploit the stacking interactions.  Helicity  can also play an important  role in the kinetics of assembly, in particular leading to frustration of bonding when strands are topologically constrained.\cite{Bois2005} 

The two length-scales also mean that single strands are ordered in a helical structure at low temperatures. At higher temperatures, where entropy dominates, they are disordered and significantly less stiff.\cite{Saenger1984,Mills1999}  Such unstacked strands are extremely flexible relative to duplexes, permitting the formation of DNA structures which involve sharply bent single-stranded regions, such as hairpins. Furthermore, stacking has significant consequences for the thermodynamics and kinetics of assembly (the role of stacking in the thermodynamics of duplex formation is discussed in Section\,\ref{FEP and fraying}). 

For complex assembly processes involving several interactions, it is  important not only to correctly reproduce properties like melting temperatures, but also the  experimentally measured transition widths so that  certain features such as hierarchical assembly are preserved. More generally, the widths of the transitions determine the response of melting temperatures to concentration changes (Section\,\ref{duplex}).  Finally,  a reasonable representation of the elastic properties  of DNA is important if the model is to be used to study systems involving DNA under stress, such as minicircles.\cite{Harris2007}

Whereas the many other top-down models in the literature each have their strengths and weaknesses, we believe that none are currently optimized for the particular suite of properties that we desire to accurately reproduce. For example, most have either ignored the stacking transition of single strands\cite{Sales-Pardo2005, Kenward2009, Ouldridge2009} or enforced helicity largely through dihedral and angular potentials imposed on the backbone of a single strand.\cite{Drukker2001, Sambriski2008, Sambriski2009, Niewieczerzal2009} In addition, where it was considered, the melting transition in previous models was generally significantly wider than experimentally reported.\cite{Kenward2009, Ouldridge2009, Sambriski2008, Sambriski2009}     In Ref.\ \onlinecite{Ouldridge_tweezers_2010} we briefly introduced a model designed to represent ssDNA, B-DNA and the transition between them, and demonstrated its utility for nanodevices by simulating a full cycle of  DNA tweezers.\cite{Yurke2000} We should note that the model is fitted at a fixed salt concentration, and does not distinguish between the strength of A-T and C-G base pairs.  

The aim of the current paper is to give a detailed description of our modeling approach.  In Section\,\ref{methods}, we present a slightly modified version of the model that appeared in Ref.\ \onlinecite{Ouldridge_tweezers_2010}, and discuss its philosophy, parameterization and simulation. The model's representation of DNA behaviour is presented in Section\,\ref{results}. We first discuss model DNA structure and thermodynamics (Sections\,\ref{basic structure}\,\&\,\ref{thermo}), before considering it's mechanical properties (Section\,\ref{mechanical}) and the representation of  certain motifs such as hairpins (Section\,\ref{structural motifs}).  Finally, we include an extensive discussion of the strengths and weaknesses of our approach in Section~\ref{discussion}.  The supporting appendices include a detailed representation of our model potential (Appendix~\ref{appendix-model}), a statistical model for stacking (Appendix~\ref{appendix-stacking}) and a statistical model for duplex formation that explicitly accounts for  the effects of stacking and fraying (Appendix~\ref{appendix-duplex}).

\section{Methods}
\label{methods}

\subsection{The model}
\subsubsection{Philosophy of the model}
\label{philosophy}

\begin{figure}
\begin{center}
\resizebox{80mm}{!}{\includegraphics{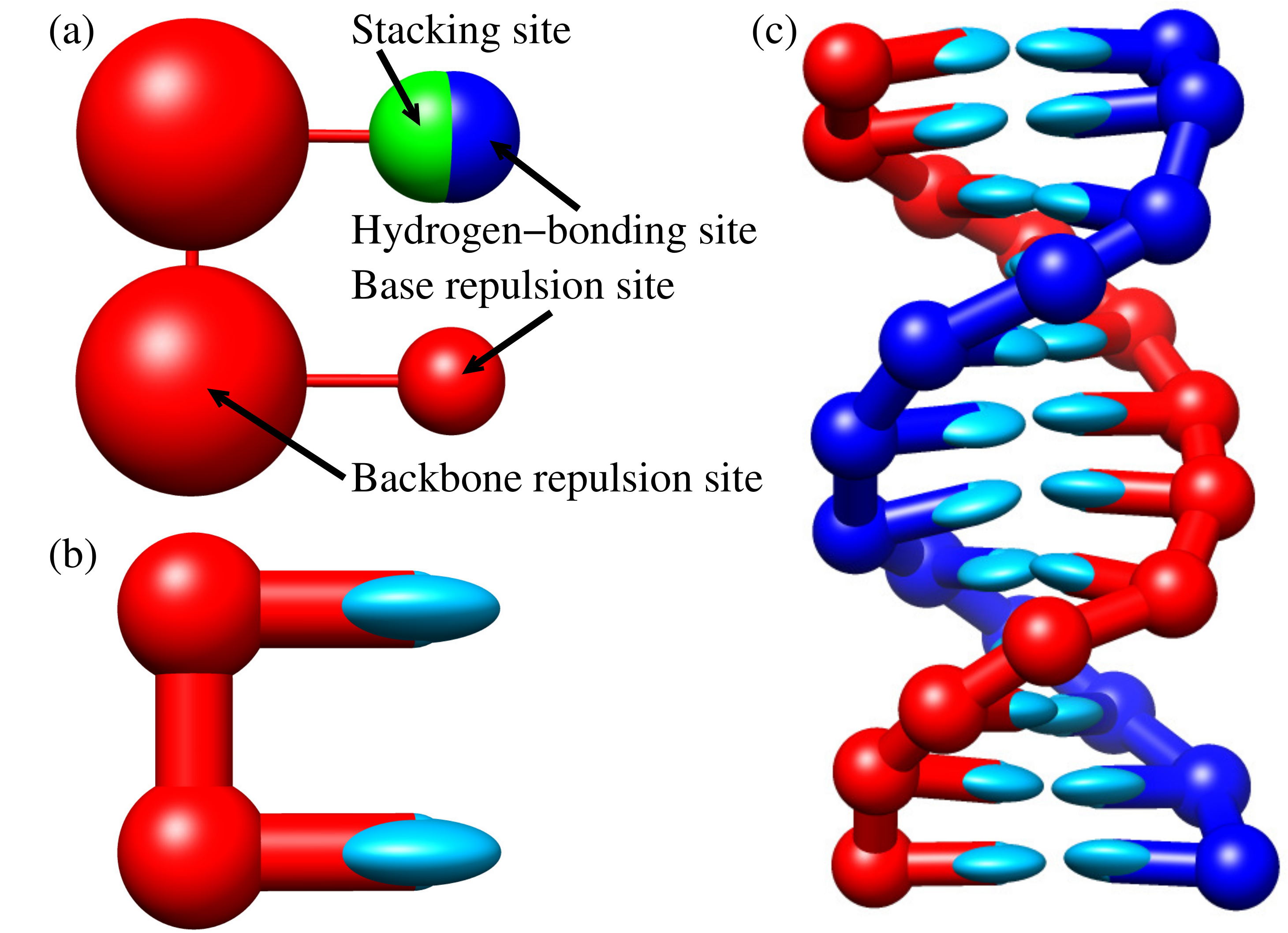}}
\end{center}
\caption{\footnotesize (a) Model interaction sites. For clarity, the stacking/hydrogen-bonding sites are shown on one nucleotide and the base excluded volume on the other. The sizes of the spheres correspond to interaction ranges: two repulsive sites interact with a Lennard-Jones $\sigma$ (Appendix\,\ref{appendix-model}) equal to the sum of the radii shown (note that the truncation and smoothing procedure extends the repulsion slightly beyond this distance (Appendix\,\ref{appendix-model})). The distance at which hydrogen-bonding and stacking interactions are at their most negative is given by the diameter of the spheres. Visualization was found to be clearer with nucleotides depicted as in (b), with the subfigures (a) and (b) representing identical nucleotides on the same scale. The ellipsoidal bases allow a representation of the planarity inherent in the model, with the shortest axis corresponding to the base normal. (c) A 12\,bp duplex as represented by the model.}
\label{model nucleotide}
\end{figure}

In designing a model, we have aimed to embed the thermodynamics of transitions involving ssDNA and dsDNA (in the most common B-form) into a 3-dimensional, dynamical, coarse-grained representation that provides a reasonable representation of structural and thermodynamic properties. This ambition naturally coincides with a top-down approach. We are not primarily concerned with the chemical details of interactions, but rather their net effect with regard to the properties of DNA.  In addition, we have attempted to capture these properties by using only pairwise excluded volume, backbone connectivity, hydrogen-bonding, stacking and cross-stacking interactions (with no explicitly length- or loop size-dependent potentials\cite{Sambriski2008, Sambriski2009, Ding2008}). The model we present here is a slightly modified version of that which appeared in Ref.\ \onlinecite{Ouldridge_tweezers_2010}, with the changes improving the representation of dsDNA flexibility and making the potential continuous and differentiable, allowing simulation methods which require forces, such as Langevin dynamics.\cite{Schlick2002}

An additional consideration in model design is the need for computational efficiency (if assembly transitions of complex structures are to be simulated). In our model, all interactions are pairwise (i.e., only involve two nucleotides, which are taken as rigid bodies). This pairwise character allows us to make efficient use of cluster-move Monte Carlo (MC) algorithms,\cite{Whitelam2009} which facilitate relaxation on all length-scales in a bound structure, and allow a much larger typical step size than possible in Langevin dynamics.

Our model consists of rigid nucleotides, illustrated in Fig.\ \ref{model nucleotide}. The three interaction sites lie in a line, with the base stacking and hydrogen-bonding/base excluded volume sites separated from the backbone excluded volume site by {$6.3$\,\AA} and {$6.8$\,\AA}, respectively. The orientation of bases is specified by a normal vector, which gives the notional plane of the base: the relative angle of base planes is used to modulate interactions (rather than through the use of off-axis sites).

\begin{figure}
\begin{center}
\resizebox{80mm}{!}{\includegraphics{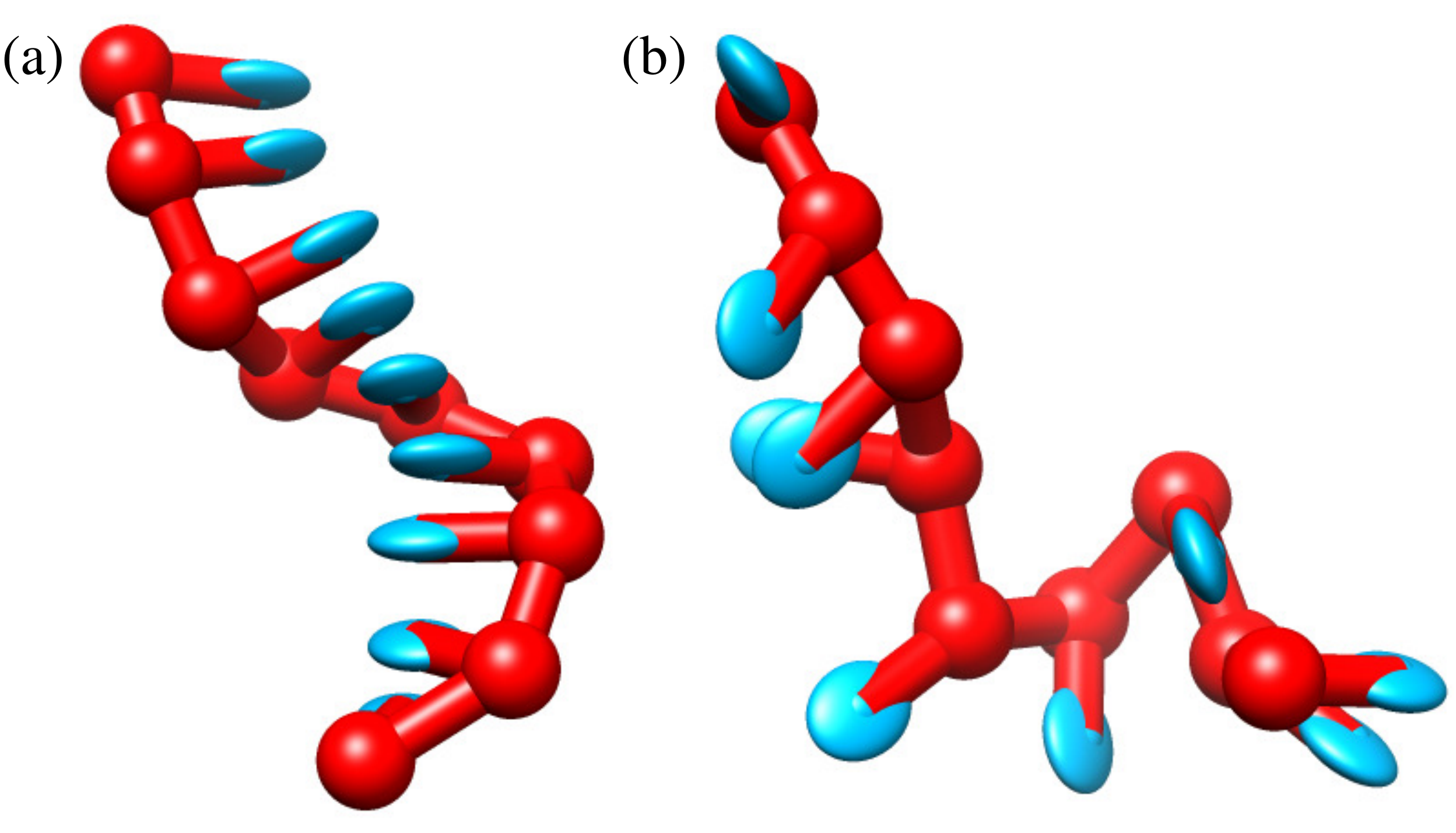}}
\end{center}
\caption{\footnotesize Two possible configurations of a 9-base strand at 333\,K. (a) All neighbours stacked to form a right-handed helix. (b) Most neighbours unstacked, giving a flexible, disordered strand.}
\label{cool pic}
\end{figure}

\subsubsection{The potential}
In this section we present an overview of the potential. Further  details are given in Appendix\,\ref{appendix-model}. Model nucleotides interact in a pairwise fashion with other nucleotides in the system. Interactions between nearest-neighbours (nn) on a strand are distinct from all others, allowing for strand connectivity and stacking. The potential can therefore be written as a sum over nn pairs, and a sum over all others: 
\begin{equation}
\begin{array}{cc}
V & =  \dSum_{\rm nn} \big(V_{backbone}  + V_{stack} + V^{\prime}_{exc}\big) \\ 
& +\dSum_{\rm other \,pairs} \big(V_{HB} + V_{c\_stack}+V_{exc}\big).
\end{array}
\end{equation}

$V_{backbone}$ is a finitely extensible non-linear elastic (FENE) spring (see Appendix\,\ref{appendix-model}), with an equilibrium length of $6.4$\,\AA, representing the covalent bonds which hold nucleotides in a strand together. 

$V_{stack}$ represents the tendency of bases to form coplanar stacks: it is a smoothly cut-off Morse potential between base-stacking sites, with a minimum at {$3.4$\,\AA}. It is modulated by angular terms which favour the alignment of normal vectors, and the alignment of the normal vectors with the vector between stacking sites. As such, the interaction encourages coplanar stacks, separated by a shorter distance than the equilibrium backbone length, leading to helical structures. Right-handed helices are imposed through an additional modulating factor which reduces the interaction to zero for increasing amounts of left-handed twist.

$V_{exc}$ and $V^{\prime}_{exc}$, representing the excluded volume of nucleotides, prevent the crossing of chains and provide stiffness to unstacked single strands.  The lack of explicit angular or dihedral potentials along the backbone allows single strands to be extremely flexible. For non-nearest neighbours, smoothly cut-off (and purely repulsive) Lennard-Jones interactions are included between all repulsion sites on the two nucleotides. For nearest neighbours, the backbone/backbone site interaction is not included because the distance between sites is regulated by the FENE spring. 

$V_{HB}$, representing the hydrogen bonds which lead to base pairing, is a smoothly cut-off Morse potential between hydrogen-bonding sites, modulated by angular terms which favour the anti-alignment of normal vectors and a co-linear alignment of all four backbone and hydrogen-bonding sites. $V_{HB}$ is set to zero unless the two bases are complementary (A-T or G-C). Together with $V_{stack}$, $V_{HB}$ causes the formation of anti-parallel, right-handed double helices for complementary strands. 

$V_{c\_stack}$ represents cross-stacking interactions between a base in a base pair and nearest-neighbour bases on the opposite strand, providing additional stabilization of the duplex.\cite{Swart2007, Sponer2006} We incorporate it through smoothed, cut-off quadratic wells, modulated by the alignment of base normals and backbone-base vectors with the separation vector in such a way that its minimum is approximately consistent with the structure of model duplexes.

Our model currently neglects some features of DNA. Although it incorporates sequence specificity (in that only A-T and C-G hydrogen bonds are possible), there is no sequence dependence in the potentials for either stacking, cross-stacking, hydrogen-bonding or excluded volume. We have made the simplifying assumption that non-complementary base pairs have zero attraction, and also neglected the possibility of alternative base-pair geometries (such as Hoogsteen\cite{Saenger1984}).We also have no explicit electrostatic interaction in the model, which may be expected to be important as bare ssDNA has a charge of $-e$ per base. For this reason, we fit to experimental data (where possible) at  ${\rm [Na^{+}]}=500$\,mM , where electrostatic properties are strongly screened.
Indeed, at these ionic concentrations, the Debye screening length is approximately 4.3\,\AA, smaller than the excluded volume diameter for backbone-backbone interactions in our model $\sim 6$\,\AA.  At the shortest distances allowed by the steric interactions, charges would have an energy of $\sim 2\,kT$ in a Debye-Huckel approximation. Other authors have attempted to explicitly include a Debye-Huckel term,\cite{Sambriski2008, Sambriski2009} but also included a salt-dependent, medium-range attraction between strands in monovalent salt to facilitate hybridization, the physical origin of which is unclear.

Many of the simplifications in our model were made to reduce the number of possible parameters. For example, sequence dependence would give 16 combinations of stacking pairs, each pair requiring several parameters to describe their interaction. We also felt that, as an initial step in modelling, it was important to obtain a good physical representation of the underlying properties of DNA assembly (such as the generic dependence of melting temperature on length), before we incorporated sequence specific or low salt effects. Furthermore, some generic effects may be obscured by sequence-specific terms (for instance, free-energy profiles such as Figure \ref{FEP} would have sequence-dependent fluctuations overlying the general trend).

\subsubsection{Parameterization of interactions}
Parameterizing such a model is a non-trivial process, as it involves a compromise between the representation of various aspects of DNA. In particular, a given parameter may influence a wide range of properties and it is difficult to design a simple metric to compare the reproduction of thermodynamic and mechanical DNA behavior. In our case, lengths were initially chosen to give our approximate B-DNA geometry. Interaction strengths and widths were then altered to give a description of the thermodynamics of stacking and duplex formation close to those in Ref.\ \onlinecite{Holbrook1999} and Ref.\ \onlinecite{SantaLucia2004}, respectively (for comparison to Ref.\ \onlinecite{SantaLucia2004}  we used an `average base pair' -- see Section\,\ref{duplex}). Finally, structural properties on long length-scales were checked, and widths of potentials and modulating factors adjusted, as potential width determines structural stiffness. Several iterations of this cycle were performed until the current parameter set was found .

In general, the interaction energy in a coarse-grained model should be interpreted as a free energy, as it incorporates a number of implicit degrees of freedom,\cite{Everaers2007} and thus it is plausible that interaction strengths could be temperature dependent. To reduce free parameters, we have avoided this temperature dependence except for the case of the the stacking strength. We found that it was difficult to design a stacking transition with an entropy as small as required (see Section\,\ref{stacking}) whilst maintaining an appropriate stiffness for dsDNA. Our stacking strength parameter has therefore been taken to be linearly dependent on temperature (see Appendix\,\ref{appendix-model}: over the range 270-370\,K, the stacking strength increases by $\sim$ 6\%), in effect reducing the entropy cost of the transition. 

There are two main possible reasons why this temperature dependence of the interaction parameters is required in our model. Firstly, it may be that it is an intrinsic property of the stacking interaction. In particular, stacking is thought to be partially a result of hydrophobic effects,\cite{Saenger1984,Guckian2000} and hence might be expected to be temperature dependent in any model without explicit water. Secondly, it may be that the coarse-graining leads to an overestimation of the entropy of the unstacked state relative to the stacked state, which then needs to be compensated by a temperature dependence in the interaction parameters.
In particular, in order to replicate the flexibility of single strands, we impose no restriction on the conformation of the backbone-backbone, backbone-base and base normal vectors except for excluded volume. This lack of constraints is certainly a significant simplification, and will allow some conformations 
that would likely be excluded by specific steric clashes in a finer-grained model (such specific geometric effects would be exceedingly difficult to reproduce in a bead-spring model such as ours). We deem this likely overestimate of available configurations to be an acceptable price to pay for the flexibility of ssDNA necessary for hairpins and nanostructures. 


\subsection{Simulation technique}
\label{simulation details}
The results reported in this paper were obtained using the Virtual-Move Monte-Carlo (VMMC) algorithm developed by Whitelam and Geissler\cite{Whitelam2009}, which allows efficient MC simulation of strongly bound systems. The algorithm takes a selected single-particle move, as with conventional MC algorithms, and then grows a cluster from connected particles according to energy changes associated with the move. The algorithm combines collective motion with the large step sizes of MC (allowing quicker decorrelation and hence equilibration). 

The combination of coarse-graining and an efficient MC algorithm provides access to processes on long time scales. To indicate simulation efficiency, we considered the formation of a 4\,bp duplex at its melting temperature, in a periodic box of side length 17\,nm (effective concentration 0.34\,mM). A recent study\cite{Kannan2009} considered a similar system using an atomistic description with continuous solvent. In the atomistic case, sophisticated sampling techniques (replica exchange molecular dynamic and umbrella sampling) were required to provide data for the transition, which was the sole focus of the study. For our model, $\sim 8$ complete binding and unbinding cycles per hour were observed for an unbiased simulation (i.e., one without enhanced sampling) performed on a single CPU core.

In order to obtain good statistics for the melting transitions, umbrella sampling\cite{Torrie1977} simulations were performed at around the melting temperature and the results extrapolated to other temperatures using single-histogram re-weighting.\cite{Kumar1992} The number of bases with negative hydrogen-bonding energy was taken as a discrete order parameter for the reaction, $Q({\bf x}^{N})$ (with ${\bf x}^{N}$ representing the coordinates of the system), and the simulations were performed using the biasing weight $\exp{(\beta W(Q)})$, with $W(Q)$ chosen iteratively to make the partial partition functions
\begin{equation}
Z^{biased}_{Q} = \int {\rm d} {\bf x}^{N} \exp \left(-\beta(V({\bf x}^{N})-W(Q^\prime({\bf x}^{N})) \right) \delta_{Q,Q^\prime} 
\end{equation}
approximately constant in $Q$. $W(Q)$ is chosen to flatten free-energy barriers, encouraging the simulation to visit rarely sampled states, thereby increasing the frequency of barrier crossing and improving statistics. We extract the unbiased partition functions using:
\begin{equation}
Z^{unbiased}_{Q} = Z^{biased}_{Q}/\exp{(\beta W(Q)}).
\end{equation}
Simulation efficiency precluded the need for multiple umbrella windows for the study of duplex formation, and the accuracy of single-histogram re-weighting was checked for 15\,bp duplexes, for which no systematic error over the range of extrapolation was found. Simulations of duplex formation were performed using two strands in a periodic box. Such simulations show strong finite-size effects due to the neglect of concentration fluctuations. These effects can be corrected for using the formalism of Ref.\ \onlinecite{Ouldridge_bulk_2010}, allowing the extraction of bulk bonding probabilities.

\section{Results}
\label{results}
\subsection{Basic structure}
\label{basic structure}
The model is specifically designed to allow an approximate representation of B-DNA in its double-stranded state. The relative sizes of the equilibrium backbone separation and ideal stacking distance lead to a pitch of 10.34 bp per turn at 296.15\,K ($23^{\rm o}$C, approximately room temperature) similar to experimental estimates of 10--10.5.\cite{Saenger1984,Sinden1994} Our model length scale is chosen so that the average rise per bp at room temperature is equal to {3.4\,\AA},\cite{Pitchiaya2006} which results in a helix with a radius (taken as the furthest extent of the excluded volume) of {11.5\,\AA}, comparable to the experimental value of {11.5--12\,\AA}.\cite{Chen2007, Pitchiaya2006}

If strands are to form a double helix, it is not possible to optimize the stacking interaction, as consecutive stacking sites cannot sit directly above one another. Single strands, however, are not constrained in this way and hence form tighter helices, with a radius approximately 80\% that of a duplex, similar to the 70-80\% observed for a number of polynucleotide single helices.\cite{Chen2007} A pleasing result is that, in order to alleviate the reduction in stacking, hydrogen-bonded bases undergo ``propellor twisting" whereby bases in a pair twist in opposite directions in order to better align their stacking centres with adjacent bases in the same strand. Experimentally, propellor twist is seen to vary from around $5^{\circ}$ to $15^{\circ}$ in GC rich regions and from $15^{\circ}$ to $25^{\circ}$ in sections with large AT content.\cite{Calladine2004} In our case we observe an average propellor twist of 21.8$^{\circ}$ at 296.15\,K, which is slightly larger than the average found for biological sequences.

\subsection{Model thermodynamics} 
\label{thermo}
\subsubsection{Single-stranded stacking transition}
\label{stacking}
The attractive stacking interaction between adjacent bases causes single strands to form helical stacks at low temperature, with this order being disrupted as the temperature increases.\cite{Saenger1984} The literature is divided on both the nature of the attraction and the thermodynamics of the transition. The relative contributions of van der Waals, induction, hydrophobic and permanent multipolar electrostatic interactions remain unclear.\cite{Guckian2000} There has also been much debate on the cooperativity with which bases stack. Vesnaver and Bresslauer claim that a 13-base strand undergoes a completely cooperative transition between helical and random coil,\cite{Vesnaver1991} whereas other authors have inferred essentially completely uncooperative transitions for the individual stacks in poly(C) and poly(A).\cite{Leng1966,Epand1967,Porschke1976, Freier1981} Other groups claim weak to moderate cooperativity, with stacking probability affected by nearby base stacking.\cite{Zhou2006, Mikulecky2006,Applequist1966} It is clear, however, that stacking has a large influence on the thermodynamics of double helix formation, as the magnitude of the enthalpy and entropy changes of hybridization increase as the single-stranded state becomes more disordered.\cite{Vesnaver1991, Holbrook1999, Mikulecky2006,  Jelesarov1999}

Given the uncertainty in stacking behaviour it is difficult to constrain the model in this regard. For simplicity we compare the model to reported uncooperative stacking (We note that introducing a large degree of cooperativity would require adding internal degrees of freedom to the nucleotide or including next-nearest-neighbour interactions). The study of Holbrook {\it et al.}\cite{Holbrook1999} is most appropriate, as it deals with heterogeneous strands rather than homopolymers, and hence might be expected to provide a reasonable estimate of the average stacking strength.

To characterize the stacking properties of our model, we simulated oligonucleotides consisting of identical nucleotides (preventing the possibility of hydrogen bonding), and recorded the distribution of the number of neighbours with a stacking interaction stronger than a minimum value \footnote{Bases were counted as stacked if their interaction was stronger than $-0.60$\,kcal\,mol$^{-1}$ (relative to a typical stacked interaction of $-6$\,kcal\,mol$^{-1}$). Adjusting the cutoff to $-1.2 $\,kcal\,mol$^{-1}$ had a negligible effect.} as a function of temperature and oligonucleotide length. For each strand length (5--9 and 14 bases), two simulations (to check convergence) were performed at $T = 333$\,K for $10^{10}$ MC simulation steps (a minimum of $7 \times 10^8$ steps per nucleotide), and we extrapolated the results to other temperatures using single-histogram reweighting. For a 14-base nucleotide, around 50\% of neighbours were found to be stacked at 338\,K, with the transition being so broad that around 30\% of neighbours remained stacked at 373\,K, and 70\% were stacked at around 306\,K.

The stacking was fitted to a simple statistical model (based on that of Poland and Scheraga for helix formation in biopolymers\cite{Poland1970}) which is discussed in detail in Appendix\,\ref{appendix-stacking}. The model contains stacking enthalpies \footnote{Our simulations are performed in the canonical ensemble, and hence should be described in terms of energy and entropy changes. We assume that, as dilute DNA strands contribute a very small partial pressure, discrepancies between constant volume and constant pressure results are small: we therefore use the term ``enthalpy'' to describe what are in fact energies in our model, for consistency with experimental literature.} and entropies $\Delta h^{st}$ and $\Delta s^{st}$, such that the statistical weight (the contribution to the partition function) of an individual pair of stacked bases is $\exp(- \Delta h^{st} / {R T} + \Delta s^{st} / {R})$ relative to the statistical weight of the unstacked state.\footnote{Throughout this article, lower case symbols represent enthalpy and entropy changes per pair of interacting bases, whereas capitals correspond to enthalpy and entropy changes per pair of interacting strands.} In addition, the statistical weight is multiplied by a cooperativity parameter $\sigma$ for each contiguous run of stacked bases, and an end-effect term $w$ for each stack which involves a base at the end of the strand. If $\sigma$ and $w$ are unity, each neighbour pair is independent. For $0 < \sigma <1$, stacking is cooperative, and for $\sigma >1$ stacking is anticooperative. For $0 < w <1$, end bases are less likely to stack, and for $w>1$ the opposite is true.

The four parameter model was fitted to data from strands of length $5-9$ bases, over a temperature range of $320-352$\,K, giving:
\begin{equation}
\begin{array}{c}
\Delta h^{st} = -5.55 \text{\,kcal\,mol}^{-1}, \\
\Delta s^{st} = -16.0 \text{\,cal\,mol}^{-1} \text{\,K}^{-1}, \\
\sigma = 0.766, \\
w = 0.783. \\
\end{array}
\label{Thermo_ss}
\end{equation}
As $\sigma$ and $w$ are close to unity, our model shows only weak cooperative and end effects. The entropy and enthalpy parameters are similar to those found by Holbrook {\it et al.},\cite{Holbrook1999} who estimated $\Delta h^{st} = -5.7$ and $-5.3$\,kcal\,mol$^{-1}$ and  $\Delta s^{st}=-16.0$ and $-15.0$\,cal\,mol$^{-1}$\,K$^{-1}$ for two different strands at [Na$^+$] = 120\,mM. Similar results at [Na$^+$] = 50mM suggest weak salt dependence in this regime.\cite{Holbrook1999}

Simulations performed in which the repulsive steric interactions were set to zero gave a slightly higher $\Delta s^{st}$ and values of $\sigma$ and $w$ consistent with unity. Thus we conclude that the small cooperative effects in our model result from excluded volume. To understand the cause of the cooperativity, consider a chain of bases $A$, $B$, and $C$, and without loss of generality, consider $B$ fixed whilst $A$ and $C$ move relative to it. Due to the requirement that base normals must point in the 3$^\prime$ to 5$^\prime$ direction to stack (see Appendix\,\ref{appendix-model}), the regions of space in which $A$ and $C$ stack with $B$ do not overlap. Therefore, if $A$ and $B$ are stacked, the excluded volume that $A$ represents to $C$ only prevents $C$ adopting conformations in which it is unstacked. By contrast, if $A$ and $B$ are unstacked, the excluded volume of $A$ can prevent $C$ adopting both stacked and unstacked configurations. As a consequence, $C$ has a slightly higher tendency to stack if $A$ and $B$ are stacked, and so there is a positive cooperativity. Similarly, end bases experience more freedom due to the reduction in excluded volume, and are therefore less likely to stack.

\begin{figure}
\begin{center}
\resizebox{80mm}{!}{\includegraphics[angle=-90]{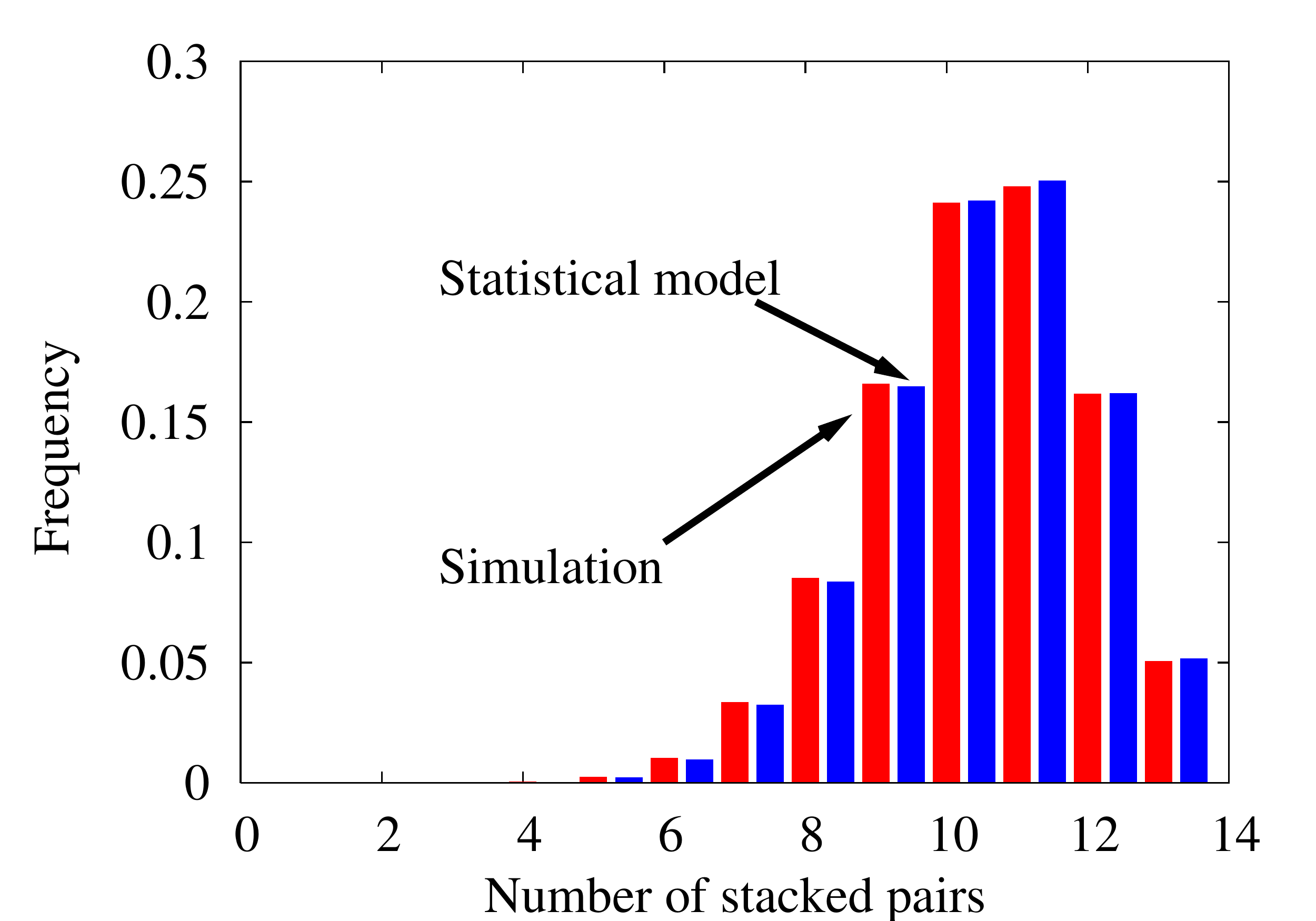}}
\end{center}
\caption{\footnotesize Frequency of the total number of stacked bases in a 14-base single strand at 300\,K from simulations of our model, and as predicted by the simpler statistical model with parameters as in Equation \ref{Thermo_ss}.}
\label{stack 14}
\end{figure}

The statistical model is very successful. Fig.\ \ref{stack 14} compares its predictions to the results for a strand length (14 bases)  and temperature (300\,K) that are both well outside the ranges that were used in the fitting.  Excellent agreement is found.

\subsubsection{Duplex formation}
\label{duplex}
Hydrogen bonding between bases can lead to the formation of bound pairs of DNA strands, which adopt the canonical `B'  double helix structure over a wide range of conditions due to stacking interactions. 
In contrast to the stacking transition, there is a reasonable consensus in the experimental literature on the melting temperature ($T_m$) of duplexes. 

We fitted our model using the two-state model and parameters of Ref.\ \onlinecite{SantaLucia2004}, which is known to give a very good prediction of experimental $T_m$. Note that we do not reproduce two-state thermodynamics (see Appendix\,\ref{appendix-duplex}), but rather treat Ref.\ \onlinecite{SantaLucia2004} as a useful parameterization of experimental results for the melting temperatures of short duplexes. As our model contains no differentiation between A-T and G-C base pairs, we compare our results to strands consisting of `average bases', the parameters for which, $\Delta h^{step}_{SL} = -8.2375 \, \rm{kcal \,mol}^{-1}$ and $\Delta s^{step}_{SL} = -22.019 \, \rm{cal \, mol}^{-1} \,\rm{K}^{-1}$, were obtained from averaging over all possible complementary base-pair steps in Ref. \ \onlinecite{SantaLucia2004}. We also use the average helix initiation terms $\Delta h^{init}_{SL} = 1.1\, \rm{kcal \,mol}^{-1}$ and $\Delta s^{init}_{SL} = 3.45 \, \rm{cal \, mol}^{-1} \,\rm{K}^{-1}$, and an additional salt correction of $\Delta s^{salt}_{SL} = -0.12754$\,cal\,mol$^{-1}\rm{K}^{-1}$ per phosphate for [Na$^+$] = 500\,mM, again taken from Ref.\ \onlinecite{SantaLucia2004}.

We simulated pairs of complementary oligonucleotides in a periodic box for a range of strand lengths between 5 and 20 bases, and extrapolated to bulk statistics using the method discussed in Ref.\ \onlinecite{Ouldridge_bulk_2010}.\footnote{Unless otherwise stated, all melting temperature calculations in this article used four simulations of $4 \times 10^{10}$ MC steps, and were performed at a reference concentration of $3.36 \times 10^{-4}$\,M. Simulations of duplexes with more than 12 bp necessitated using a larger periodic cell, and hence a lower concentration. The fraction of bound duplexes was scaled to the higher concentration assuming the separate species are approximately ideal, as justified in Ref.\ \onlinecite{Ouldridge_bulk_2010}.} Umbrella sampling, using the number of base pairs with a negative hydrogen-bonding energy as an order parameter $Q$, was used to ensure good sampling.

For the purposes of comparison with Ref.\ \onlinecite{SantaLucia2004}, we defined a state to be bound if any hydrogen-bonding interaction between strands had an energy below a cutoff of  $-0.60\,\rm{kcal \,mol}^{-1}$, with typical hydrogen-bonding energies of a single base pair being larger by a factor of approximately 7. Doubling the cutoff had no significant effect on our results. $T_m$ was taken as the temperature at which half of the strands would be bound in a bulk solution.

\begin{figure}
\begin{center}
\resizebox{85mm}{!}{\includegraphics{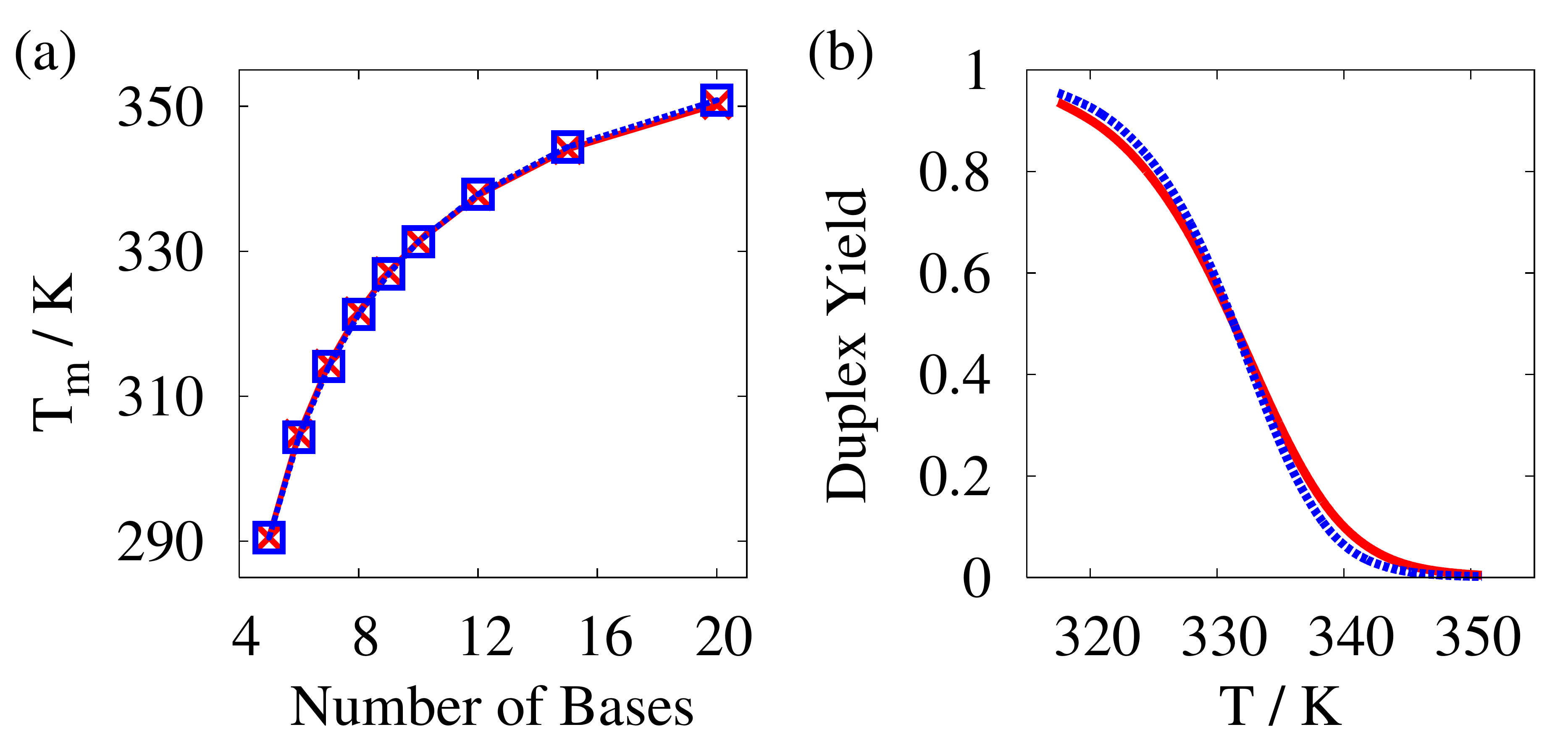}}
\end{center}
\caption{\footnotesize (a) $T_m$ as a function of strand length at an equal strand concentration of $3.36\times 10^{-4}$\,M, as given by our model (crosses connected by a solid line) and averaged parameters from Ref.\ \onlinecite{SantaLucia2004} (squares connected by a dashed line). (b) Fraction of 10-base strands bound in duplexes at a concentration of $3.36 \times 10^{-4}$\,M as a function of temperature, from our model (dashed line) and using the parameters of Ref.\ \onlinecite{SantaLucia2004} (solid line).}
\label{Tm_duplex_transition}
\end{figure}

The variation in melting temperature with duplex length is shown in Fig.\ \ref{Tm_duplex_transition}\,(a), where it is compared to the predictions of the model of Ref.\ \onlinecite{SantaLucia2004}. The agreement in the dependence of $T_m$ on length is extremely good: this dependence is essentially a measure of the cooperativity of the duplex forming transition, which is most strongly influenced by the relative contributions of hydrogen-bonding and stacking/cross-stacking to duplex stability.

The polynucleotide melting temperature (the melting temperature for infinitely long strands) at {500\,mM} $[\rm{Na}^+]$ for a strand of 50\% C-G content, is predicted by the empirical relations given by Blake and Delcourt\cite{Blake1998} and Frank-Kamenetskii\cite{Frank-Kamenetskii1971} as {365.8\,K} and {363.2\,K}, respectively. An approximate value for our model can be estimated by simulating a pair of long, complementary strands in a partially bound state, and finding the temperature at which the free-energy change of adding an additional base pair to a partially formed duplex is zero. Simulations of partially formed 100--bp strands (with the duplex/single-stranded DNA interface at a variety of points) gave values of $T$ in the range 364--366\,K, in good agreement with the empirical relations.

Fig.\ \ref{Tm_duplex_transition}\,(b) compares the 10-bp duplex yield as a function of temperature for our model with the predictions of Ref.\ \onlinecite{SantaLucia2004}. The widths of the transitions are consistent to within a few degrees Kelvin, with our model consistently producing a marginally sharper transition for all duplex lengths. The width of the transition determines the response of the system to changes in concentration. Consider, for example, a simple two-state model of DNA hybridization, as used in Ref.\ \onlinecite{SantaLucia2004} and expressed in Eqn.\ (\ref{2sm}). Assuming equal total concentrations of each strand ($[A_0]$), the width of the transition scales approximately as:
\begin{equation}
\Delta T \sim \frac{k_{\rm B}T_m^2}{\Delta H},
\end{equation}
and the change in $T_m$ with concentration is given by:
\begin{equation}
\frac{dT_m}{d[A_0]} = -\frac{k_{\rm B}T_m^2}{[A_0] \Delta H} \sim \frac{\Delta T}{[A_0]},
\end{equation}
and hence agreement in both $T_m$ and the transition width at a given concentration imply agreement in $T_m$ over a range of concentrations.

\subsubsection{Free energy profile of duplex formation and fraying}
\label{FEP and fraying}
\begin{figure}
\begin{center}
\resizebox{80mm}{!}{\includegraphics[angle=-90]{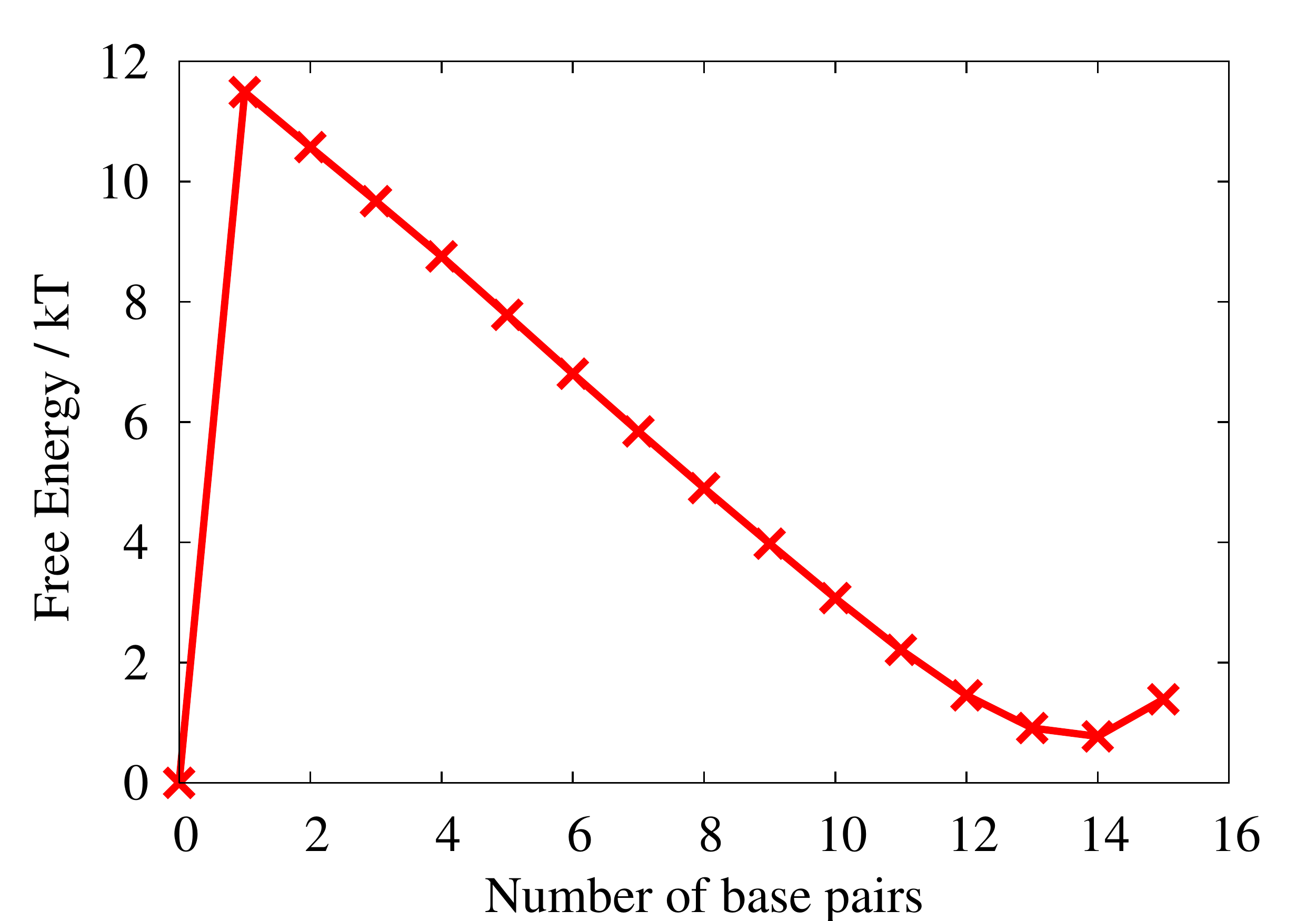}}
\end{center}
\caption{\footnotesize Free-energy profile of bonding of a 15 bp duplex, as a function of the number of base pairs, at 343\,K.}
\label{FEP}
\end{figure}
The free energy of duplex formation of a 15-bp duplex is plotted in Fig.\ \ref{FEP} as a function of the number of base pairs (the order parameter for our umbrella sampling). To avoid complicating features in the free-energy profile due to hairpins and misbonds, which can conceal the underlying trends at low numbers of bonds, only base pairs that are present in the desired duplex had a non-zero strength of hydrogen bonding in this simulation. The general form of the free-energy profile is qualitatively similar to that found in Ref.\ \onlinecite{Sambriski2009} for another coarse-grained model of DNA, with an initial entropy penalty for the formation of the first base pair, followed by a downhill slope as the duplex `zips up' in a cooperative fashion. As can be seen, the formation of the final  base pair is actually free-energetically unfavourable, and the typical state consists of a duplex with `frayed' ends. This fraying arises because bases at the end of the duplex lack the stabilizing influence of neighbouring base pairs on either side and entropy favours the open state.

Although fraying is a widely accepted phenomenon,\cite{Andreatta2006} experimental data is rather sparse, though it is established that weaker AT ends fray more easily than CG capped helices.\cite{Nonin1995} Nonin {\it et al.}\cite{Nonin1995} inferred fraying probabilities of terminal AT bps of around 0.375 and 0.7 at {273\,K} and {298\,K} respectively, and found 0.015 and 0.12 for GC pairs at the same temperatures (at moderate salt concentrations), whereas Patel {\it et al.}\cite{Patel1975} found much higher melting temperatures for terminal base AT pairs, concluding that they were around 50\% frayed at {313\,K} at high salt concentration. Our model shows approximately 10\% fraying at {273\,K}, increasing to around 21\% at {300\,K} and reaching 50\% at approximately {330\,K}, reasonable values for `average' base pairs. We note that in many cases, particularly at low temperature, end bps in our model break but remain stacked, adopting conformations to maximize stacking at the expense of hydrogen bonding.

\subsubsection{Effect of stacking and fraying on thermodynamics of duplex formation}
\label{dH(T)}
We attempted to fit the duplex yield as a function of temperature, for each strand length $l$, using a two-state model of the form in Eqn.\ \ref{2sm}. 
\begin{equation}
\frac{[A_l B_l]}{[A_l][B_l]} = v \frac{Z_{ll}}{Z_l^2} = \exp\Big(-\beta \big(\Delta H_l - T \Delta S_l\big)\Big),
\label{2smb}
\end{equation}
where $[A_l]$ is the concentration of strand $A$ of length $l$ and $[B_l]$ and $[A_l B_l]$ are the concentrations of its complementary strand and the bound pair. $v$ is the volume simulated, $Z_{ll}$ and $Z_l$ are the statistical weights (contributions to the partition function) of duplexes and single strands of length $l$ in our simulations and $\Delta H_l$ and $\Delta S_l$ the (assumed $T$-independent) enthalpy and entropy of transition (we note that for our simulations in the canonical ensemble, $\Delta H$ corresponds to the change in internal energy of the system).
It was found, however, to be an unsatisfying fit to the melting curves, and further attempts to fit $\Delta H_l$  and $\Delta S_l$ as a linear function in $l$ (by analogy with the nearest-neighbour model), were unsuccessful. This failure should not come as a surprise, however, as several authors have indicated that the entropy and enthalpy of duplex formation show temperature dependence due to the single-stranded stacking transition.\cite{Vesnaver1991, Holbrook1999, Mikulecky2006,  Jelesarov1999} A more sophisticated model which explicitly treats the stacking and fraying is detailed in Appendix\,\ref{appendix-duplex}. We show that, for our model, temperature dependent effects can be incorporated into an extended nearest-neighbour description of the transition.

The actual temperature-dependent transition enthalpy can be deduced from:
\begin{equation}
\Delta H= -\frac{\rm d}{\rm d \beta} \ln K_{eq},
\end{equation}
where $K_{eq}$ is the equilibrium constant of the reaction. 
The enthalpy changes at $T_m$ for our model are slightly larger than predicted by Ref.\ \onlinecite{SantaLucia2004}, which is to be expected as the transitions are slightly narrower. The discrepancy rises from about 6\% for  5-bp duplexes to around 22\% for 20-bp double strands. The behaviour of $\Delta S$ is similar.

\begin{figure}
\begin{center}
\resizebox{80mm}{!}{\includegraphics[angle=-90]{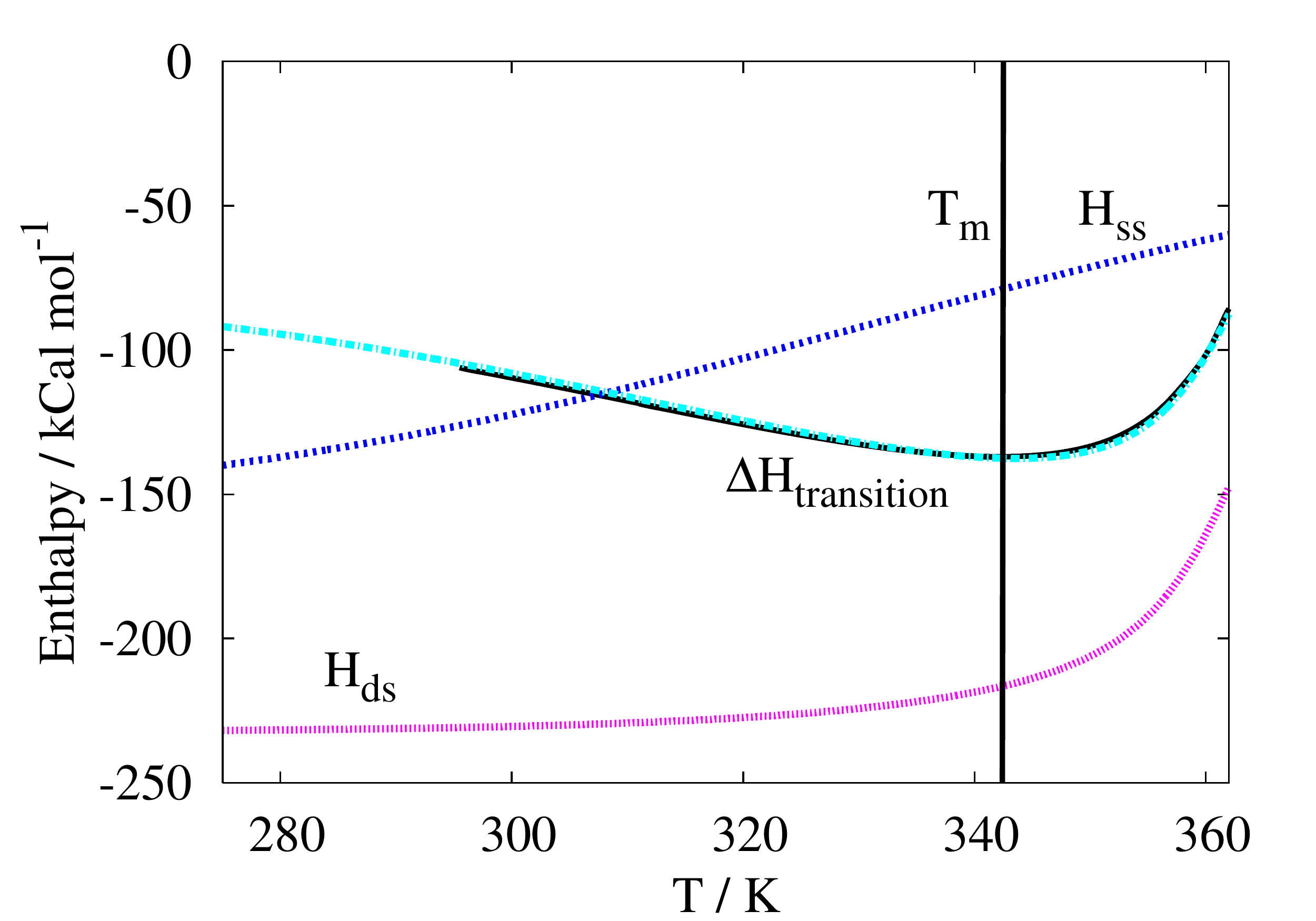}}
\end{center}
\caption{\footnotesize Variation with $T$ of enthalpies associated with the formation of a 15-bp duplex. Solid lines represent simulation results, dashed lines the predictions of the statistical model outlined in Appendix\,\ref{appendix-duplex}. 
The lines labeled $\Delta H_{\rm transition}$ give the enthalpy change upon duplex formation for the simulations and the statistical model. The lines labeled $H_{\rm ds}$ and $H_{\rm ss}$ are the enthalpies of the duplex and single strands respectively, relative to a completely unstacked state. The transition enthalpy in the statistical model is the difference between the latter two curves. The vertical line denotes the melting temperature $T_m=342.6 \,K$.}
\label{dH_T}
\end{figure}

To investigate the details of the temperature dependence of enthalpy changes in duplex formation, we simulated 15\,bp duplex formation over a wide range of temperatures (for clarity, we again only give ``correct" pairs an attractive hydrogen-bonding interaction), with the data shown in Fig.\ \ref{dH_T}. We find that at low temperatures (up to $342$\,K) $\Delta H$ becomes more negative with increasing temperature, with a gradient that reaches a maximum size of around $-0.055$\,kcal\,mol$^{-1}$\,K$^{-1}$ per base pair at approximately 328\,K. At $342$\,K, however, $\Delta H$ reaches its most negative value, before increasing rapidly towards zero for higher temperatures.

The statistical model of Appendix\,\ref{appendix-duplex} allows us to analyze this behaviour in terms of the enthalpy changes within the bound and unbound states.
As shown in Fig.\ \ref{dH_T}, the enthalpy of the bound state is  approximately constant at lower temperatures, whereas the enthalpy of the single strands becomes less negative with increased temperature as they unstack, causing the observed tendency for $\Delta H$ of the transition to become more negative with increasing temperatures.  As temperature continues to increase, however, the typical bound state changes from being a fully-formed duplex at low temperatures to a higher enthalpy partially-melted state at higher temperatures. Thus the enthalpy of the bound state becomes less negative as fraying becomes more significant, resulting in the observed increase in $\Delta H$.

This change in enthalpy due to the stacking transition has been observed experimentally by several groups,\cite{Holbrook1999, Mikulecky2006, Tikhomirova2004, Vesnaver1991, Jelesarov1999} who deduced values for the typical enthalpy gradient of $-0.050$, $-0.05$ to $-0.1$, $-0.062$, $-0.095$  and $-0.068$ to $-0.87$\,kcal\,mol$^{-1}$\,K$^{-1}$ per base pair, respectively, in reasonable agreement with our model. These investigations were generally performed with either oligonucleotides with several CG pairs at the end \cite{Holbrook1999, Mikulecky2006, Vesnaver1991, Jelesarov1999} or polynucleotides,\cite{Tikhomirova2004} both of which would massively reduce the impact of fraying. If we set the fraying contribution to zero, we obtain a typical value of $-0.06$ to $-0.07$\,kcal\,mol$^{-1}$\,K$^{-1}$, in even better agreement with experiment.

In addition, Jelesarov {\it et.\,al.} \cite{Jelesarov1999} also considered a duplex with AT bps at the end of the helix, for which $\Delta H$ becomes more negative with increasing $T$ at low temperature, before flattening-off by around {310 K}, in agreement with the predictions of our model for the consequences of fraying. Measurements were not performed at high enough $T$ to check for an eventual reversal of the gradient of  $\Delta H$ with temperature, but our model predicts the effect should be observable. In particular, duplexes with large AT end regions and stabilizing GC cores should demonstrate such an effect.

\subsection{Mechanical properties}
\label{mechanical}

\begin{figure}
\begin{center}
\resizebox{80mm}{!}{\includegraphics{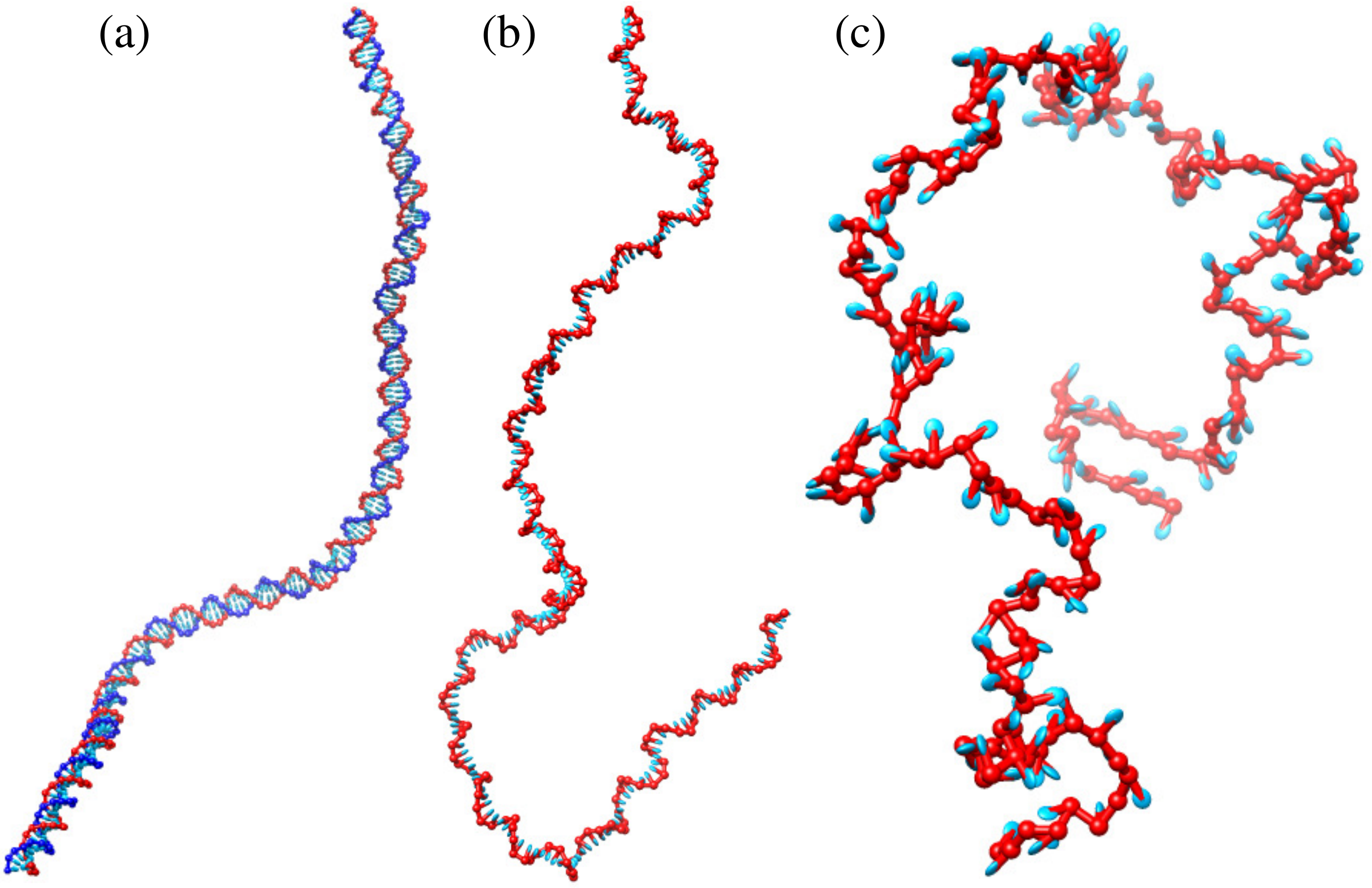}}
\end{center}
\caption{\footnotesize Typical configurations indicating relative flexibility of double-stranded, stacked single-stranded and unstacked single-stranded DNA. (a) 202\,bp double helix at 296.15\,K. (b) Stacked single strand of 202 bases at 277.15\,K. (c) Unstacked single strand of 160 bases at 296.15\,K.}
\label{p_l}
\end{figure}

\subsubsection{Single-stranded persistence length}
Single strands, particularly when unstacked, are extremely flexible relative to dsDNA. This is crucial for nanotechnology, as it allows structures to contain highly bent ssDNA regions, such as at the vertices of polyhedra or the hinges of nanomachines. 

Poly(dT) (long single stands of DNA in which all the bases are thymine) is generally assumed to be entirely unstacked at room temperature, and has little tendency to form secondary structure.\cite{Mills1999, Saenger1984} As a consequence, it can be used to test the inherent flexibility of unstacked single strands. Gapped helices have been used by Mills {\it et al.},\cite{Mills1999} who inferred a high salt persistence length of {20--30\,\AA} from rotational decay rates, and Rivetti {\it et al.},\cite{Rivetti1998} who studied length distributions with atomic force microscopy, finding $\sim 16${\,\AA} for short sections ($<5$ bases), growing to around {28\,\AA} for longer regions. Fluorescence resonance energy transfer between donors and acceptors attached to either end of poly(dT) has also been used to fit polymer models to chain end-to-end distributions, with Murphy {\it et al.} finding a persistence length of around {19.4\,\AA} at 500\,mM [Na$^+$].\cite{murphy2004} All of these results suggest persistence lengths on the scale of 2-5 bases.

\begin{figure}
\begin{center}
\resizebox{80mm}{!}{\includegraphics[angle=-90]{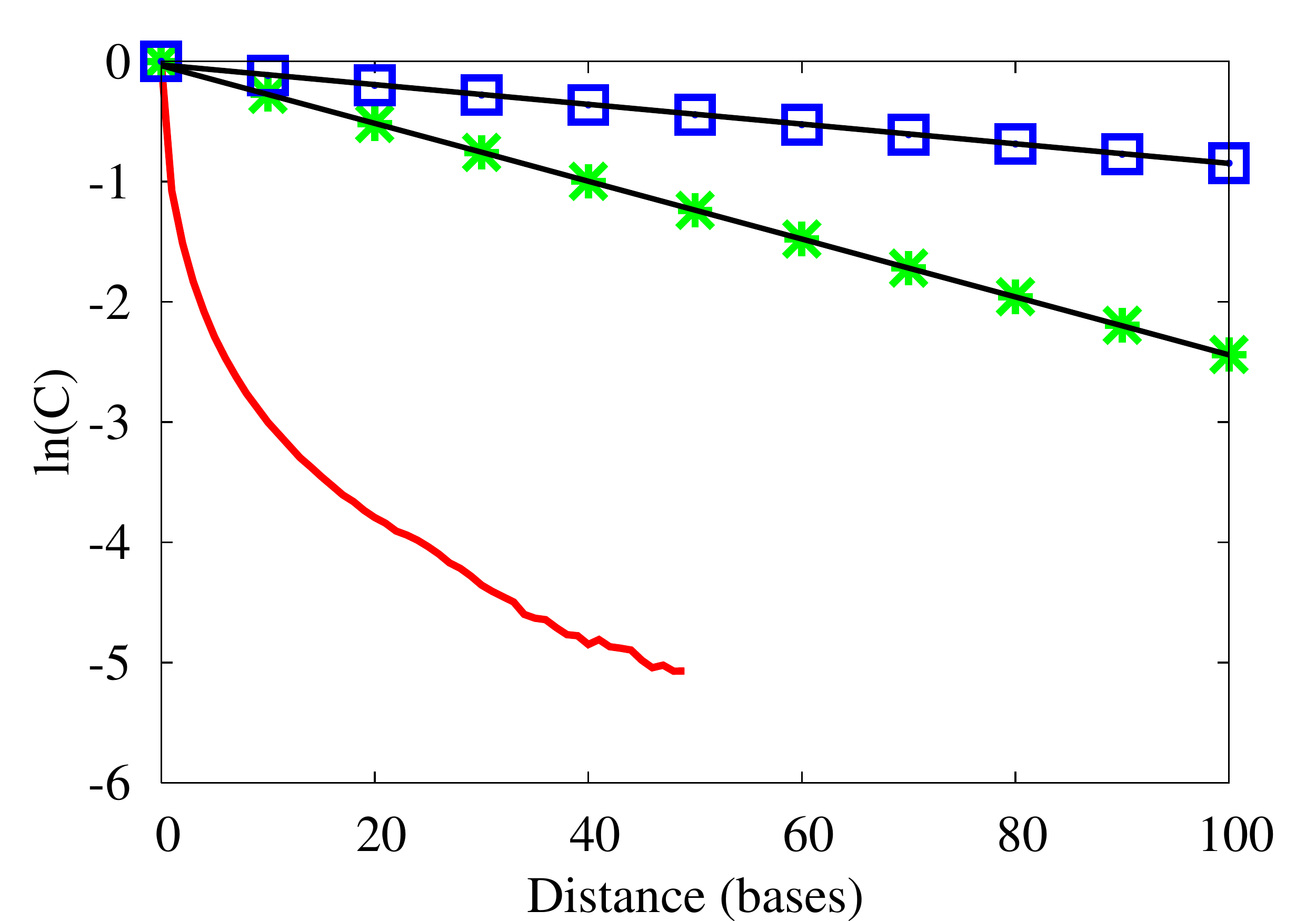}}
\end{center}
\caption{\footnotesize Decay of the correlation ($C$) of helix axis plotted against base separation for a duplex at 296.15\,K (squares) and a stacked single strand at 277.15\,K (stars). The lines are fits to exponential decays. Also shown (solid line, no symbols) is the decay of the correlation of backbone vectors for an unstacked single strand at 296.15\,K.}
\label{pl_stacked}
\end{figure}

To compare our model to experiment, we simulated single strands of one base type with stacking interactions set to zero, as shown in Fig.\ \ref{p_l}\,(c), to mimic poly(dT). Poly(dT) is sometimes modeled as a worm-like chain,\cite{murphy2004,Rivetti1998} in which a local stiffness opposes bending, resulting in an exponential decay of the correlations of backbone vectors with distance. In our model, however, unstacked ssDNA is essentially a freely-jointed chain with excluded volume, meaning that the conformation of backbone sites is restricted by steric clashes rather than local stiffness. As a result, the correlation of backbone-backbone vectors decays slower than exponentially (Fig.\ \ref{pl_stacked}), due to steric interactions between non-neighbouring nucleotides. Such a decay implies that adjacent bases demonstrate larger kinking than would be expected from the picture of a worm-like chain with an equivalent overall stiffness of the strand. Consecutive backbone orientation is restricted only by steric clashes, hence large kinks are possible. More distant bases, however, still feel the excluded volume, and so the tendency is for directional correlation to decay slowly. 

\begin{figure}
\begin{center}
\resizebox{80mm}{!}{\includegraphics[angle=-90]{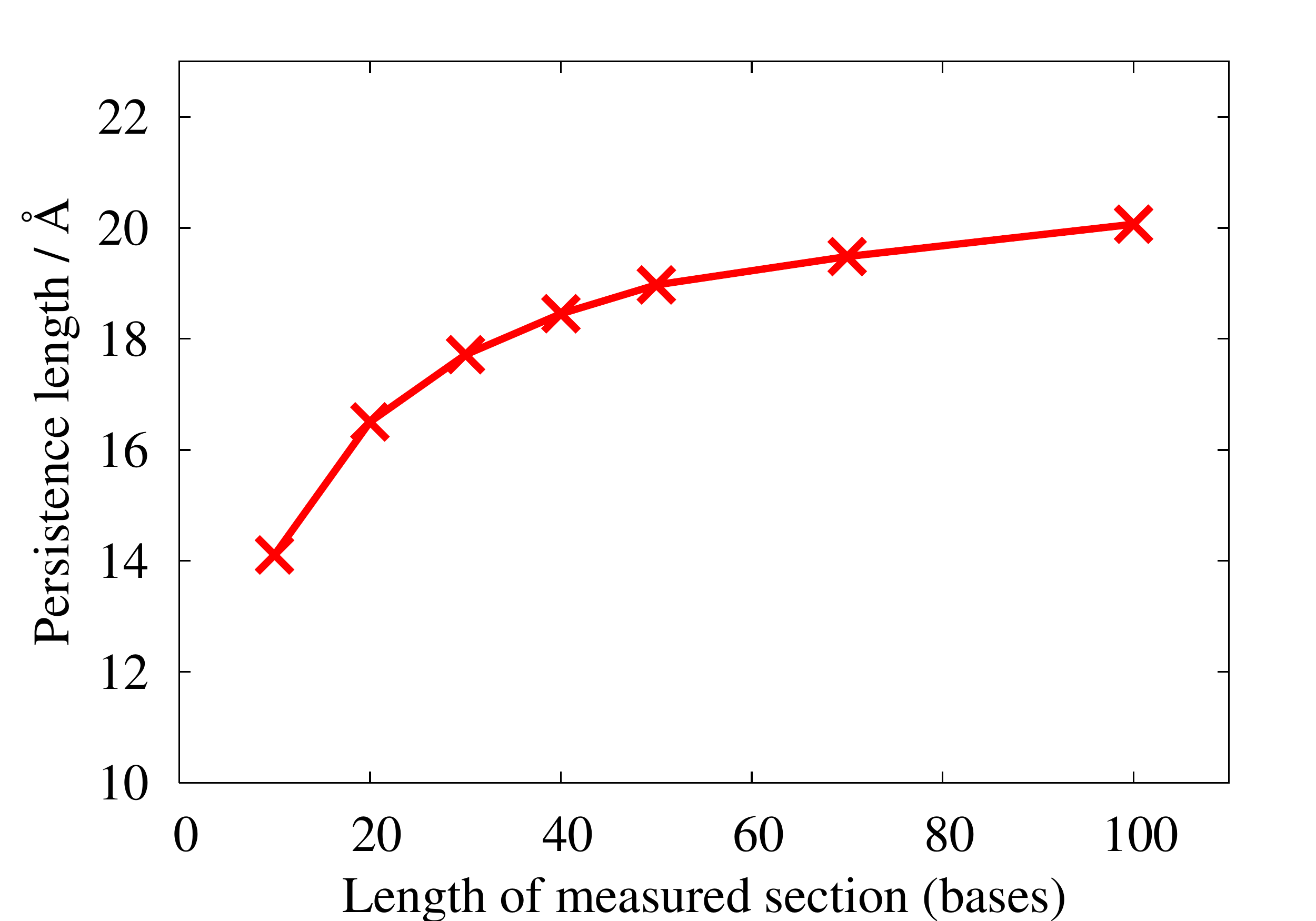}}
\end{center}
\caption{\footnotesize $L_{ps}$ plotted against the length of the single-stranded region of DNA analyzed  at 296.15\,K. For the purposes of comparison, the separation of successive backbone sites is approximately $6.4$\,\AA.}
\label{ss_pl_unstacked}
\end{figure}

It was therefore difficult to obtain an unambiguous value for the persistence length to compare to experiment. We used the general definition from Ref.\ \onlinecite{Cifra2004}:
\begin{equation}
L_{ps} = \frac{\langle{\bf L} . {\bf l_0}\rangle}{\langle l_0 \rangle},
\label{persistence length}
\end{equation}
with ${\bf L}$ being the end to end vector of the strand and ${\bf l_0}$ representing the first backbone-backbone vector. As the strand approaches infinite contour length, the value of $L_{ps}$ should tend towards a constant, $L_{ps}^{\infty}$. We estimated $L_{ps}^{\infty}$ by evaluating Eqn.\ \ref{persistence length} for single-stranded regions of lengths from 10 to 100 bases, embedded within strands of 70 to 160 bases. Four simulations were performed at {296.15\,K} for each length for at least $2.5 \times 10^8$ MC steps per particle, with the results plotted in fig.\,\ref{ss_pl_unstacked}. The reason for embedding the measured length in a longer strand is that bases toward the end of single- or double-stranded DNA possess an increased relative flexibility. In order to obtain persistence length values that are valid for long strands where end effects are negligible bases near the end of strands were ignored.

This increased flexibility at the ends, which results from fewer restraining interactions, may be manifested in experimental systems. It is possible that interpretations that rely on the configuration of bases at the end of strands may be biased by such increased flexibility. 
Our results show that for strands of ${\sim100}$ bases, the persistence length is similar to experimentally inferred values ($19-30$\,\AA). $L_{ps}$ continues to grow noticeably for contour lengths much larger than $L_{ps}^{\infty}$ ($\sim 20$\,\AA, or just over three bases), an effect that is consistent with the findings of  Rivetti {\it et al.},\cite{Rivetti1998}, and which indicates that non-nearest-neighbour interactions are important in providing the effective stiffness.   This agreement with experimental results, together with the fact that our model provides a reasonably good representation of the effective excluded volume, suggests that our prediction that the freely-jointed chain gives a better representation of the conformational statistics of unstacked single-stranded DNA than the worm-like chain picture, should also hold for real DNA.

Mills {\it et al.}\cite{Mills1999} also investigated the flexibility of gapped duplexes connected by poly(dA) at $4^{\circ}$C, when the bases are largely stacked into single helices. Although the interpretation depends on the probability of stacking, the intrinsic persistence length of the stacked regions was estimated to be in the region of {100\,\AA}, corresponding to approximately 30 bases (stacked regions have a shorter length per segment than unstacked sections due to twisting). This value is noticeably larger than that for unstacked strands, but smaller than for duplexes (approximately 150 bases at high salt concentration). For comparison we simulated single strands of 202 identical bases at $4^{\circ}$C for  $8 \times 10^9$ MC steps (ignoring the data from the five bases at either end), requiring that all bases maintained a stacking interaction of {$\ge -0.60$\,kcal\,mol$^{-1}$}  with their neighbours (doubling this value had no discernible effect). Unlike in the unstacked case, excluded volume does not play a large role as the length scale over which bending occurs is much larger than the size of one base (as can be seen in Fig\ \ref{p_l}\,(b)). Hence, the relative alignment of vectors between stacking sites (which now act as the basic steps along the strand) was observed to decay exponentially, allowing a fit of the form:
\begin{equation}
\langle {\bf l_n}.{\bf l_0} \rangle= \exp ({-n  \langle l_0 \rangle /L^{stack}_{ps}}),
\label{persistence_length_stacked}
\end{equation}
from which we concluded that $L_{ps}^{stack}/\langle l_0 \rangle = 41.5$ bases (see Fig.\ \ref{pl_stacked}) for our model. This value is higher than that reported by Mills {\it et al.}\,\cite{Mills1999}  by approximately 50\%, but importantly it is much greater than the persistence length of unstacked ssDNA whilst also being much more flexible than dsDNA. Furthermore, the estimates in Ref.\ \onlinecite{Mills1999} assume unstacked bases behave as regions of persistence length {30\,\AA}, which is at the upper end of estimates for poly(dT). As already noted, our results suggest that local kinking can be much larger than would be implied by the persistence length of unstacked bases. As such, the flexibility contribution from a single unstacked base may be larger than estimated, and consequently the flexibility of the stacked regions may be overestimated, possibly bringing our model into better agreement with the data.

\subsubsection{Double-stranded persistence length}
The persistence length of dsDNA is generally accepted to be approximately {450-500\,nm} at moderate to high [Na$^+$], corresponding to around 130--150 base pairs.\cite{Hagerman1988, Baumann1997} We performed three simulations of a duplex of length 202\,bp at {296.15\,K} for $1.5 \times 10^9$ MC steps, ignoring the data from the ten base pairs at either end. Similar to our findings for stacked single helices, the correlation of the helix axis  (defined as the distance between consecutive base-pair midpoints) at two points was observed to decay exponentially with distance, allowing an estimate of $L_{ps}^{duplex}$ through Eqn.\ \ref{persistence_length_stacked}. Fig.\ \ref{pl_stacked} indicates a model persistence length of around 125 base pairs, in reasonable agreement with experiment. A typical configuration is shown in Fig.\ \ref{p_l}\,(a).

\subsubsection{Double-stranded torsional and extensional stiffness}
Torsional rigidity (in the linear regime) is quantified by an elastic modulus $C$, which relates applied torque $G$ to resultant twist $\Delta \theta$ of a duplex of length $l$: $C = G l/ \Delta \theta$. Estimates for $C$ have been made using cyclization kinetics and topoisomer distributions for minicircles,\cite{Hagerman1988, Crothers1992, Vologodskaia2002} luminescence depolarization \cite{Fujimoto2006} and from twisting of DNA under tension,\cite{Bryant2003} giving values in the range {170--440\,fJ\,fm}. The effect of salt concentration on $C$ is not entirely clear from the experimental literature.\cite{Fujimoto2006}

Calculating the response to torsion is non-trivial, as the curvature of the DNA axis makes the twist between two ends hard to define. In our previous work,\cite{Ouldridge_tweezers_2010} we attempted to infer an elastic modulus from the fluctuations in the angle between successive bases when projected onto the plane perpendicular to the vector joining their midpoints. Unfortunately, this method overestimates the torsional flexibility, presumably failing to decouple torsional variation from other fluctuations in a base-pair step. In this work, we instead obtain an approximate estimate of the torsional modulus by considering the twisting of the central 10 base pairs of a 20\,bp duplex, and the central 20 base pairs of a 30\,bp duplex at $296.15$\,K. Such short sections are extremely stiff, minimizing the natural bending fluctuations. To provide an unambiguous definition of torsion and twist, MC moves were chosen so that the base pairs at the end of the central section remained perpendicular to the vector between their midpoints, allowing the vector between the midpoints to define an axis about which torsion could be applied and twist measured. 

Simulations were performed in which the torque applied to the end bases was varied between $\pm 8$\,pN\,nm, and the resultant twist used to infer $C$. A separate estimate was also obtained using the equipartition result for the variance in twist at zero torque: $\langle \Delta \theta_{\rm twist}^2 \rangle = k T l /C$. Further simulations used the equipartition result to estimate $C$ under a tension of 9\,pN, to ensure that stretching the duplexes had no effect. All estimates (for both 10- and 20-bp regions of interest) gave $C \sim 455 - 495$\,fJ\,fm, suggesting that this is a reasonably robust estimate of the torsional stiffness of DNA duplexes in our model. 

A long molecule of dsDNA under low tension responds as an extensible worm-like chain, with the behaviour initially dominated by the straightening of the chain, before stretching the base-pair rise itself becomes relevant as the chain extension approaches the contour length.\cite{Wang1997, Wenner2002} At higher forces, the duplex undergoes an overstretching transition and the B-DNA structure breaks down.\cite{Smith1996} Experimental estimates for the extensional modulus $K$, obtained from fitting force-extension curves to extensible worm-like chain models, give $K$ in the region of 1050--1250\,pN at high salt.\cite{Wang1997, Wenner2002}

\begin{figure}
\begin{center}
\resizebox{80mm}{!}{\includegraphics[angle=-90]{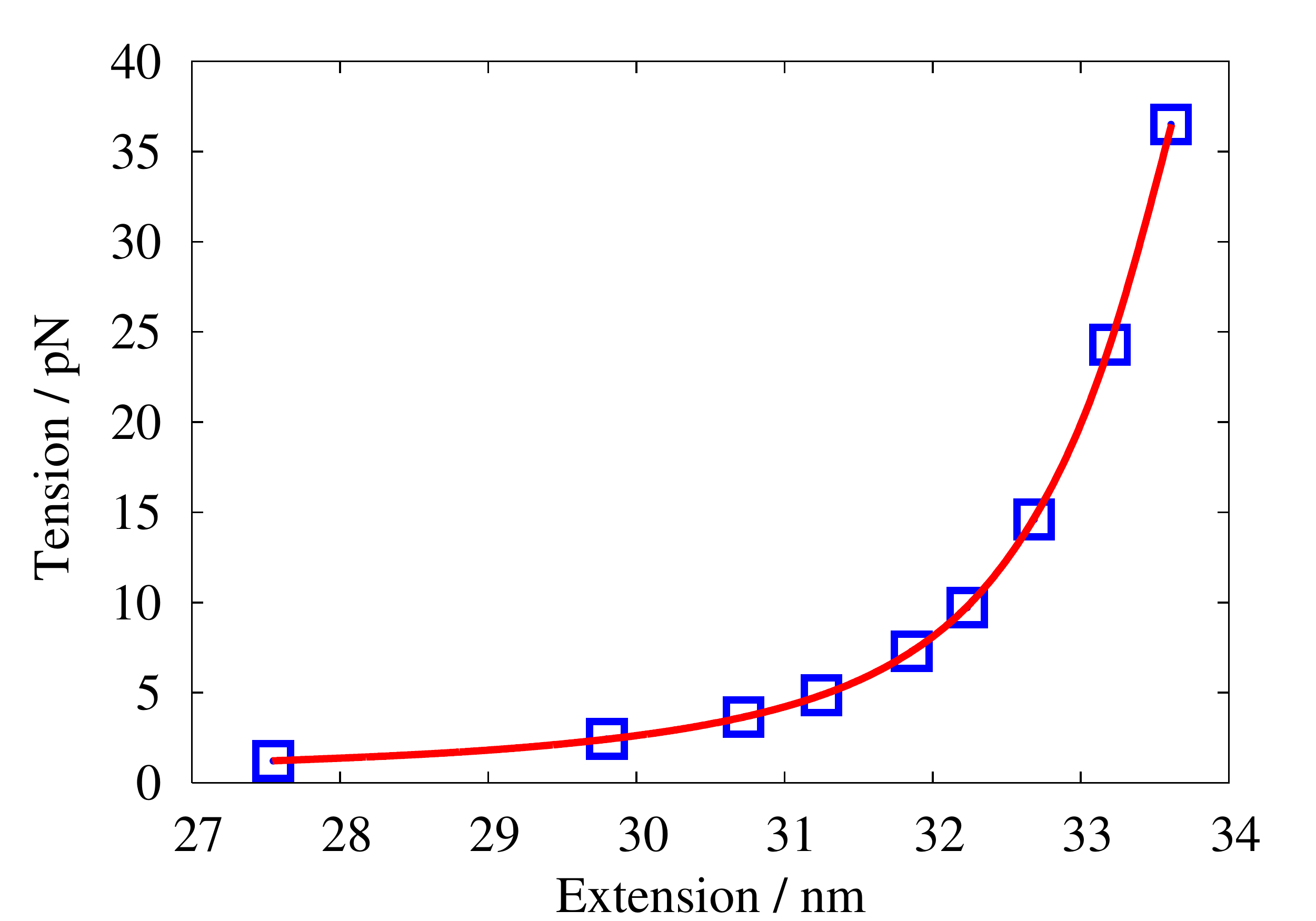}}
\end{center}
\caption{\footnotesize Tension applied against extension for the central 100 bp of a 110-bp duplex at 296.15\,K. The squares are simulation results, the solid line is a fit using Eqn.\ \ref{Odijk}.}
\label{force-extension}
\end{figure}

The extensional modulus $K$ was estimated by applying tension to a 100-bp region within a 110-bp double helix, and fitting the resultant force-extension curve to the result of Odijk\cite{Odijk1995} for extensible worm-like chains:
\begin{equation}
x = L_0\left(1+\frac{F L_0}{K} - \frac{k T}{2 F}\left[1 + y \coth y \right] \right),
\label{Odijk}
\end{equation}
where
\begin{equation}
y= \left(\frac{F L_0^2}{L_{ps} k T}\right)^{1/2},
\end{equation}
in which $x$ is the extension resulting from a force $F$ applied to a duplex of contour length $L_0$ and persistence length $L_{ps}$. Performing an unconstrained three-parameter fit with the values of $L_0$, $L_{ps}$ and $K$ gave an excellent agreement with the data, as shown in Fig.\ \ref{force-extension}, with $K=2120$\,pN, {$L_0=339.4$\,\AA} and {$L_{ps} = 438$\,\AA} (129\,bp). The value of $L_0$ is similar to that expected from the rise of a short duplex ({exactly 3.4\,\AA} per base pair would give $L_0=336.6$\,\AA), and $L_{ps}$ is only slightly larger than the estimate from the decay of the correlation of the helix axis (415\,\AA). This agreement suggests that the extensible worm-like chain model provides a good description of the model's properties in this regime, and that the value of $K=2120$\,pN is a reasonably robust one for our model.

Our model gives $C \approx 475$\,fJ\,fm (slightly larger than the top of the experimental range of {170--440\,fJ\,fm}) and $K \approx 2120$\,pN, (about twice as large as typical experimental estimates). We do not believe the differences are crucial to the processes we are interested in investigating (although certain quantities, such as the critical twist density at which plectonomes are extruded, will be affected). It was found to be difficult to reparameterize the model to reduce these moduli without decreasing the persistence length, which is already slightly below experimental estimates. We feel that the current compromise, in which the persistence length is most faithfully reproduced, is a reasonable one as it is easier to imagine that nanostructures and nanodevices would be more sensitive to bending than torsional or extensional stiffness.

It is worth noting that recent investigations have suggested that DNA overwinds when stretched.\cite{Gore2006} Our model does not reproduce this anti-intuitive behaviour, instead slightly untwisting as the stacking distance is extended. It is possible, therefore, that the model fails to capture the softness of a mode of deformation -- perhaps the sloping of base pairs with respect to the axis\cite{Lionnet2006} --that leads to this behaviour. If this is the case, it is perhaps unsurprising that the estimated moduli are larger than experimental observations

\subsection{Structural motifs}
\label{structural motifs}
\subsubsection{Hairpins}
\begin{figure}
\begin{center}
\resizebox{80mm}{!}{\includegraphics{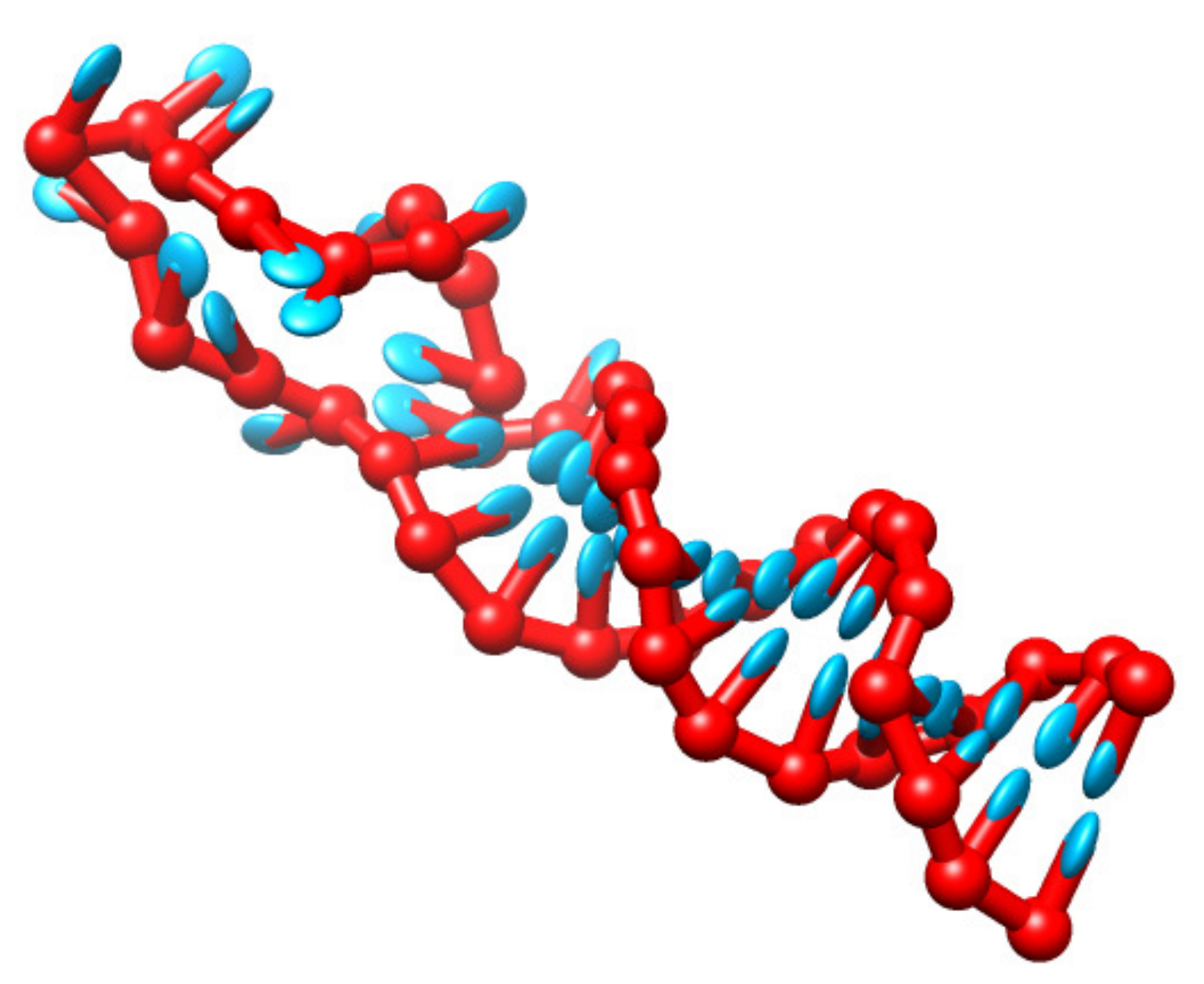}}
\end{center}
\caption{\footnotesize A hairpin with a 12\,bp stem and an 18-base loop at 343\,K.}
\label{hairpin}
\end{figure}

DNA hairpins, which occur when a self-complementary strand binds to itself and forms a duplex stem and an unhybridized loop (Fig.\ \ref{hairpin}), are a common structural motif. They have biological importance as a mechanism for release of superhelicity through cruciform formation.\cite{Sinden1994} Their relevance to nanotechnology includes metastable states (either occurring by accident\cite{Ouldridge_tweezers_2010} or through design).\cite{Bois2005, Green2008} In addition, they are an extremely common motif in biological RNA structures.\cite{Hendrix2005} Aside from our earlier work using a previous parameterization of the current model,\cite{Ouldridge_tweezers_2010} we are unaware of any  simultaneous application of a coarse-grained model to the formation of both hairpins and bimolecular duplexes. Our approach, in which the single strands have the potential to be extremely flexible, allows for hairpins and duplexes to have appropriate relative stabilities. 

\begin{figure}
\begin{center}
\resizebox{80mm}{!}{\includegraphics{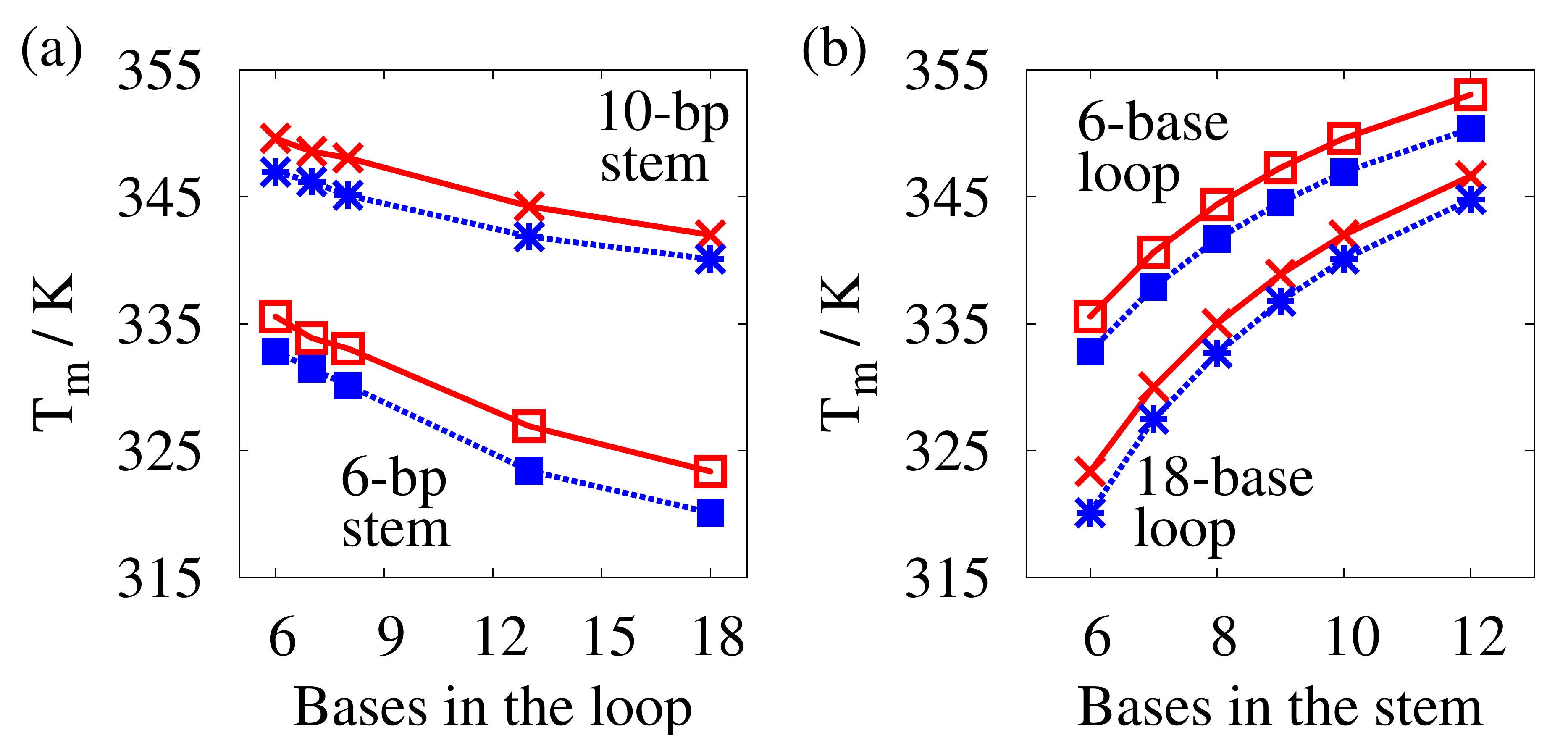}}
\end{center}
\caption{\footnotesize Variation of hairpin melting temperature with (a) loop length and (b) stem length from our model (symbols connected by dashed lines) and from Ref.\ \onlinecite{SantaLucia2004}. 
}
\label{Tm_loop_stem}
\end{figure}

To demonstrate the ability of our model to represent hairpins, we simulated systems with stem sizes ranging from 6--12 bps, and loops of 6--18 bases. Four simulations for each hairpin were performed in the vicinity of $T_m$ for $4 \times 10^{10}$ MC steps (corresponding to at least $10^9$ steps per nucleotide). Umbrella sampling as a function of hydrogen-bonded base pairs was used to ensure good statistics. In this case, we considered only states with at least one of the `native' bps in the stem present as being a hairpin, as long loops have the potential to form transient base pairs with little relevance to the stability of the target structure. SantaLucia has presented parameters for estimating the melting temperature of hairpins,\cite{SantaLucia2004} which we again take as a good representation of experimental results. These parameters include sequence independent entropy penalties for loop formation and enthalpy/entropy terms for the stabilizing effect of the first mismatched bp in the loop (called a `terminal mismatch': we compare to an average $\Delta h^{term}_{SL} = -2.91 \, \rm{kcal \,mol}^{-1}$ and $\Delta s^{term}_{SL} = -7.33 \, \rm{cal \, mol}^{-1} \,\rm{K}^{-1}$). Our results for $T_m$ are compared to the predictions of Ref.\ \onlinecite{SantaLucia2004} in Figs.\ \ref{Tm_loop_stem}\,(a)\,and\,(b). $T_m$ is defined as the temperature at which a strand is in a hairpin state half of the time. 

The results indicate that our model slightly underestimates $T_m$ for hairpins relative to the predictions of Ref.\ \onlinecite{SantaLucia2004} (and by extension, experiment) by approximately {3\,K}, which is slightly less than 1\% of the absolute melting temperature (at the $T_m$ predicted by Ref.\ \onlinecite{SantaLucia2004}, our hairpins constitute approximately 25\% of the ensemble rather than 50\%).  Encouragingly, the trends with loop length and stem size are well reflected by our model (this is particularly pleasing, as the dependence on loop length was not used in parameterization), an indication that the majority of the physics of hairpin formation is well represented by our model. We note that our model is less successful for the smallest loops (3-5 bases), possibly because it does not incorporate specific interactions within a tightly packed loop that may provide extra stability.\cite{Kuznetsov2001} As found with duplex formation, transition widths for our model are slightly smaller than predicted by Ref.\ \onlinecite{SantaLucia2004} (the difference is very similar to that observed in Fig.\ \ref{Tm_duplex_transition}(b)).

\subsubsection{Mismatches, bulges and internal bubbles}
\label{mismatches etc}
A variety of other DNA motifs exist, such as duplexes involving mismatches between non-complementary base pairs or with one strand carrying extra, unpaired bases. SantaLucia\,\cite{SantaLucia2004} has provided parameters for the influence of these motifs on $T_m$. In many cases, they are highly sequence dependent and it is less clear than in the simple double helix case (where the variations in parameters are relatively smaller) that averaging over $\Delta S$ and $\Delta H$ contributions for all sequences is a reasonable approach to find an average effect. It should, however, give a rough estimate of the typical change in melting temperature due to a motif. 

\begin{table*}

\begin{center}
\begin{tabular}{c  c  c  c  c}
\hline
\hline
& & & & \\
{\bf Motif}  &{\bf Complementary bp} &  {\bf Motif size} & \multicolumn{2}{c}{  {\bf $\Delta T_m$} / K }\\
&  & & {\bf Our Model}  & {\bf Ref.\ \onlinecite{SantaLucia2004}} \\ \hline 
& & & & \\
Dangling end & 5 & 1 base &  $+3.95$ & $+4.24$ \\
& 8 & 1 base & $+1.20$ & $+1.44$  \\
& 15 & 1 base & $+0.74$ & $+0.61$  \\ \hline
& & & & \\
Bulge & 8 & 1 base & $-18.58$ & $-23.40$ \\
& & 2 bases & $-24.64$ & $-27.23$ \\
& 15 & 1 base & $-8.86$ & $-12.58$ \\
&  & 2 bases & $-11.51$ & $-11.67$ \\
&  & 5 bases & $-16.91$ & $-13.78$ \\ \hline
& & & & \\
Terminal mismatch & 5 & 1 base / strand &  $+6.85$ & $+6.95$ \\
& 8 & 1 base / strand & $+2.73$ & $+2.55$  \\
& 15 & 1 base / strand & $+0.74$ & $+0.63$ \\ \hline
& & & & \\
Internal mismatch & 8 & 1 base / strand & $-8.77$ & $-14.09$ \\
/ bubble & & 2 bases  / strand & $-15.77$ & $-21.86$ \\
 & & 5 bases / strand & $-25.83$ & $-28.81$ \\ 
& 15 & 1 base / strand & $-5.35$ & $-4.97$ \\
&  & 2 bases / strand & $-9.53$ & $-11.60$ \\
&  & 5 bases / strand & $-15.62$ & $-15.74$ \\ 
\hline
\hline
\end{tabular}
\end{center}
\caption{\footnotesize Effect on the melting temperature of a complementary duplex due to the addition of a motif. In this table, $\Delta T_m$ is the difference between the $T_m$ of a structure with the motif and a fully complementary duplex consisting of the same number of complementary bps as the motif structure. For internal mismatches, bulges and bubbles, the motif was placed at the centre of the duplex.}
\label{motifs}
\end{table*}

We compared the effect of several motifs on model duplex $T_m$  to the predictions of Ref.\ \onlinecite{SantaLucia2004}, again averaged over all possible sequences (Table \ref{motifs}). 
The simplest possible case is that of a single unpaired base at the end of a strand, generally referred to as a `dangling end'. Typically, dangling ends are observed to provide a stabilizing influence, assumed to result from cross-stacking with the final base pair of the duplex, although the degree of stabilization is highly sequence dependent.\cite{SantaLucia2004, Guckian2000} The cross-stacking interaction included in our model provides such a stabilizing effect, and the degree of stabilization is in good agreement with the predictions of Ref.\ \onlinecite{SantaLucia2004}. 

In contrast to dangling ends, extra, unpaired bases on one strand within the helix are highly destabilizing, as they disrupt the helix structure. In the terminology of SantaLucia, these are known as bulges. In general, our  model slightly underestimates the destabilization of helices due to bulges compared to the predictions of Ref.\ \onlinecite{SantaLucia2004}, although the observed  melting temperatures remain within 2\% of the predictions.

If a non-complementary pair of bases is added to an otherwise complementary duplex to form a mismatch, the effect is generally stabilizing at the end of a duplex (this is a ``terminal mismatch") and destabilizing in the interior. Our model reproduces this tendency as shown in Table\,\ref{motifs}, and also captures the increase in destabilization if the mismatch region is extended (to form an internal ``bubble"). Once again, the destabilizing effect of motifs internal to the duplex tend to be slightly underestimated relative to the predictions of Ref.\ \onlinecite{SantaLucia2004}, and the observed melting temperatures again remain within around 2\% of the predictions.

The motifs provide a good test of the model, as many were not considered in parameterization (although the dangling ends and terminal mismatches were used to constrain the strength of cross-stacking). In addition, misbonded structures involving these motifs may have a role in the kinetics of nanostructure assembly, and hence it is important that the model provides a reasonable representation of them.
Although in some cases the quantitative agreement with Ref.\ \onlinecite{SantaLucia2004} is not perfect, the model represents these motifs in a physically sensible way and the trends in stability at least qualitatively reflect the average properties of DNA. Furthermore, the typical magnitudes of $\Delta T_m$ are reasonable, with the $T_m$ remaining within 2\% of the average predictions of Ref.\ \onlinecite{SantaLucia2004}. It is possible that an underestimate of the disruptive effect of extra bases on the helical structure,\cite{Sinden1994} perhaps because the excluded volume of bases is smaller than in reality, causes the underestimate of $\Delta T_m$ due to internal motifs. This effect, however, would be expected to be larger for bulges than for mismatched pairs or symmetric bubbles.

\begin{figure}
\begin{center}
\resizebox{80mm}{!}{\includegraphics[angle=-90]{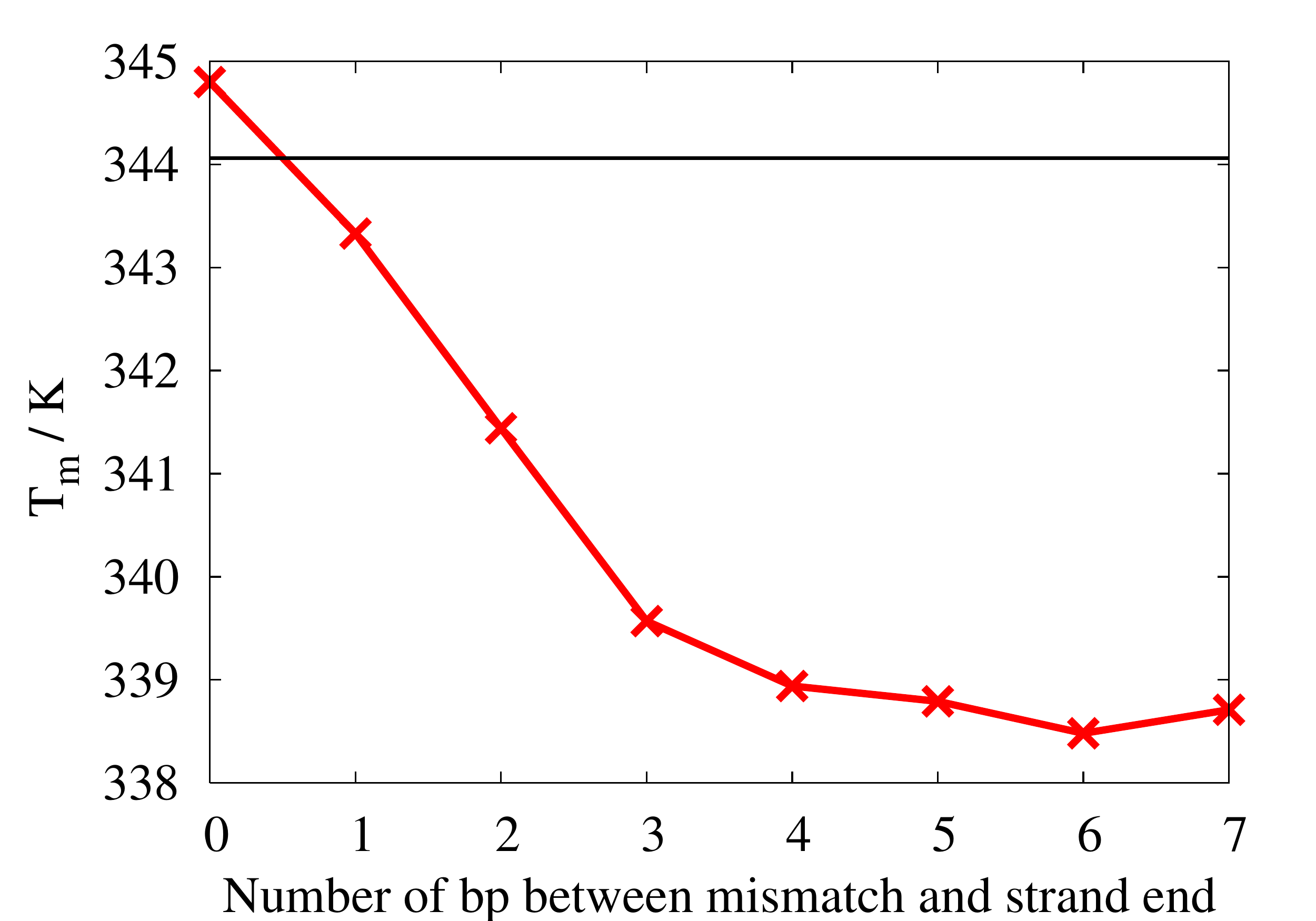}}
\end{center}
\caption{\footnotesize Melting temperature of 15-bp complementary helix with an additional mismatch added against the distance of that mismatch from the end of the strand. The melting temperature in the absence of a mismatch is indicated via the horizontal line.}
\label{Tm_mismatch}
\end{figure}

Given the good agreement between the model and Ref.\ \onlinecite{SantaLucia2004} for a single mismatch added to a 15-bp duplex, we investigated how the position of the  mismatch affected stability. $T_m$ is plotted against the position of the mismatch in Fig.\ \ref{Tm_mismatch}. As can be seen, there are two distinct regimes, with the melting temperature initially decreasing as the mismatch is moved from the end of the strand (where it is stabilizing) towards the centre. Eventually, however, it reaches a plateau at around five bases from the end of the strand. 

\begin{figure}
\begin{center}
\resizebox{80mm}{!}{\includegraphics[angle=-90]{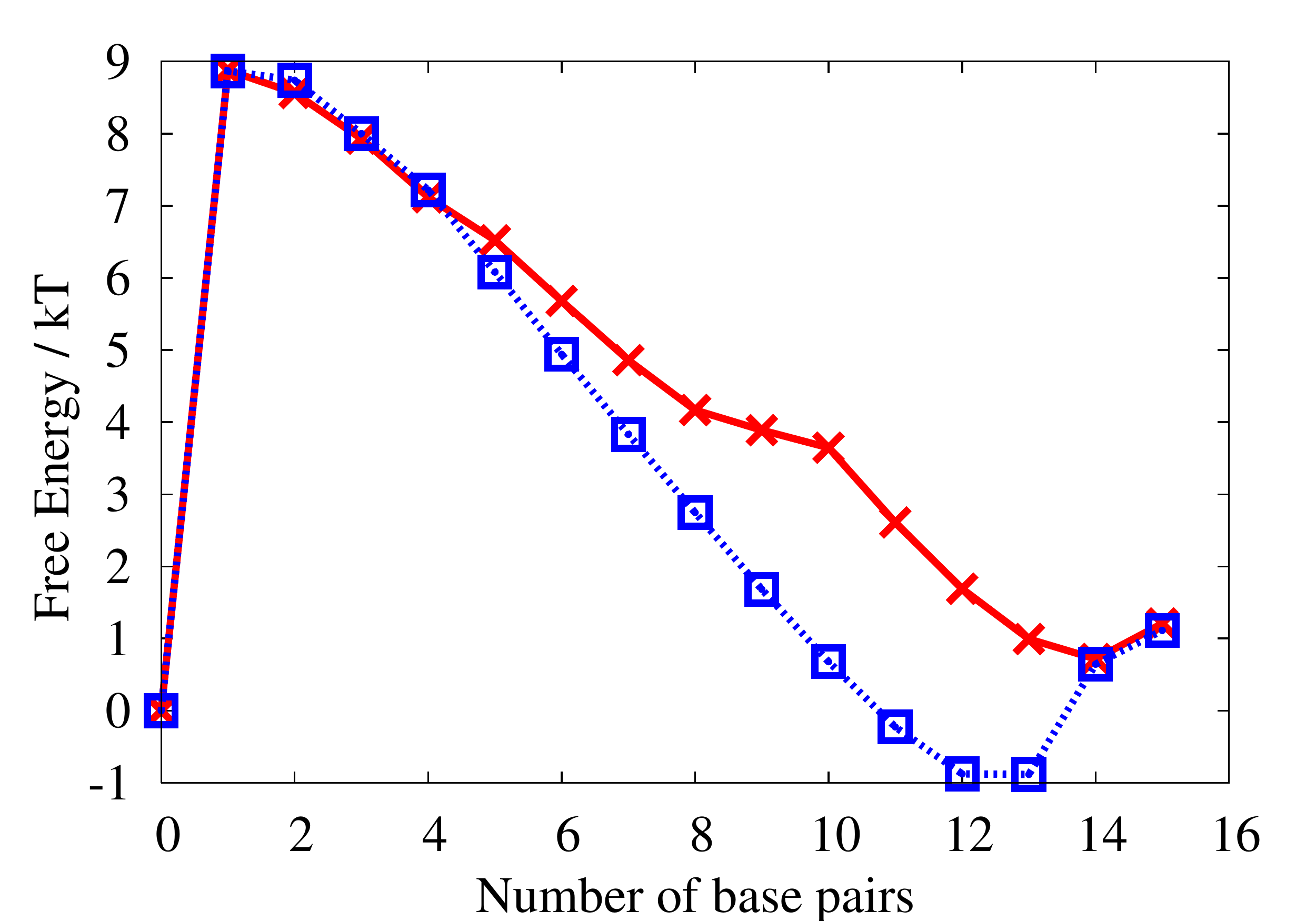}}
\end{center}
\caption{\footnotesize Free energy profile at 339\,K for a 15 base pair duplex with one additional mismatch placed 2 bases from the end (squares) and 6 bases from the end (crosses).}
\label{fep_mismatch}
\end{figure}

The cause of this plateau can be identified from examining the free-energy profiles for duplexes with mismatches located two and six bp from the end (Fig.\ \ref{fep_mismatch}). The first point to note is that the stability of duplexes with the maximum number of base pairs (15) is nearly identical, despite the difference in mismatch position. This suggests that provided a mismatch is surrounded by base pairs on either side, changing its location has little effect on the total free energy. The difference in $T_m$ arises instead from a difference in the lowest free-energy state.

When the mismatch is near to the strand end (in the regime where $T_m$ depends on mismatch position), the most stable state consists of the larger section of duplex formed with the bases beyond the mismatch unpaired. In this regime, the total free-energy gain from pairing the bases beyond the mismatch does not compensate for the free-energy cost of enclosing a mismatch in a helix. As the mismatch is moved towards the centre, the larger section loses bases and so becomes less stable, with the consequence that $T_m$ drops. At some point, however, it becomes favorable for the bases in the shorter region to also bond. From this point onwards, the most stable state consists of the two duplex regions surrounding the mismatch.  The net effect of moving the mismatch further towards the centre only marginally affects the overall stability of the duplex.  As a result a plateau in $T_m$ should occur. 

As the temperature is lowered, the free-energy gain from base pair formation increases. As a consequence, the number of bases required before the region beyond the mismatch is stable as a duplex decreases. For example, we find that for a mismatch two bases from the end of a 15 bp duplex, the enclosed mismatch state becomes the most stable just below 320\,K.

\begin{figure}
\begin{center}
\resizebox{80mm}{!}{\includegraphics{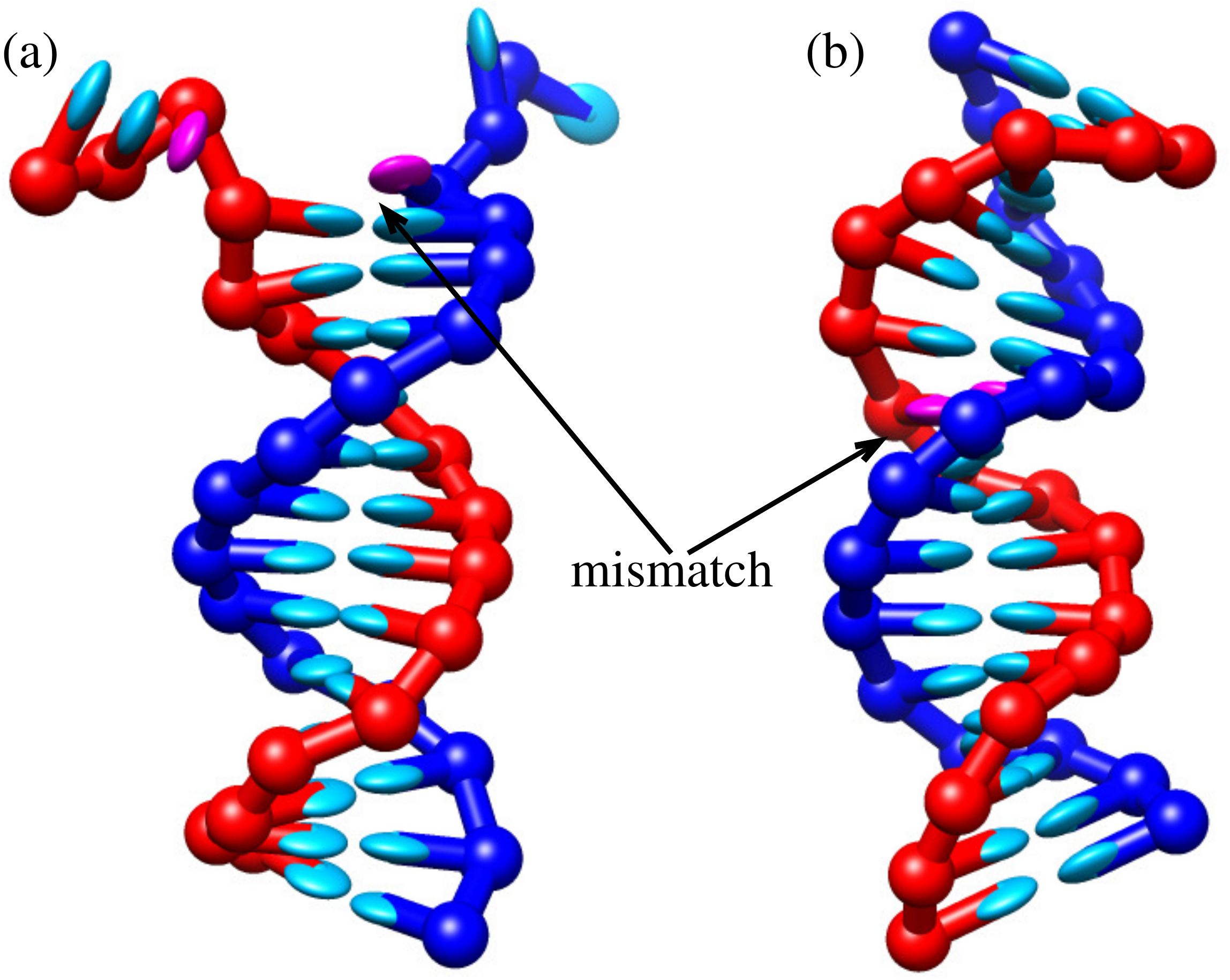}}
\end{center}
\caption{\footnotesize Typical configurations of a duplex with 15 complementary bp and 1 internal mismatch at 335\,K. a) Mismatch two bp from the end of the strands, with unpaired bases after the mismatch. b) Mismatch six bp from the end of the strand, enclosed by two intact helices. }
\label{cool pic 2}
\end{figure}

It is claimed in Ref.\ \onlinecite{SantaLucia2004} that the stability of a mismatch is independent of its position, except for terminal mismatches and mismatches occurring one base from the end, which may cause the final base pair to be unstable. Our simulations suggest, however, that the distance of the mismatch from the duplex end at which $T_m$ plateaus should increase with strand length (as longer strands melt at higher temperature). Furthermore, a similar temperature-dependent influence of motif location should hold for all destabilizing internal bubbles and bulges, as the beginning of the plateau simply indicates the point at which it is free-energetically favourable to enclose the disruption. This result should be qualitatively robust to the approximations in the model. In particular, sequence dependence will likely cause fluctuations but not destroy the general trend.


\section{Discussion}
\label{discussion}
We have examined in detail the structural, mechanical and thermodynamic properties of a coarse-grained model of DNA based on that presented in Ref.\ \onlinecite{Ouldridge_tweezers_2010} (and used there to simulate a full cycle of DNA tweezers, an iconic nanodevice).  Several small alterations to the model  were made in order to improve the description of DNA flexibility and allow for the calculation of forces and torques. The aim of the model is to embed the known thermodynamics of B-DNA into a dynamical, coarse-grained representation of DNA while simultaneously providing a reasonably accurate description of the structural and mechanical properties of B-DNA and ssDNA. 

The model provides a good quantitative representation of the three key thermodynamic processes that affect self-assembly: single-stranded stacking, duplex hybridization and hairpin formation. To our knowledge, this is the first coarse-grained model for which all three processes have been considered simultaneously.

The mechanical properties of DNA are also reasonably well represented by the model, with the singled-stranded persistence length (for stacked and unstacked bases) and double-stranded persistence length, stretch modulus and torsional modulus all of similar size to typical experimental estimates. Importantly, the inclusion of the stacking transition allows single strands to be unstacked and flexible, which facilitates the formation of hairpins as well as other DNA nanostructures for which single-stranded regions are important.

The model contains several simplifications, the most important of which are the lack of sequence dependence beyond the specificity of A-T and G-C bonds and the absence of explicit electrostatic interactions. Thus the model cannot predict screening effects without a new parameterization at each salt concentration. Furthermore, in its current state, the model may incorrectly represent structures that involve the close proximity of strands that are not bound to each other, where it would fail to capture the cumulative repulsion resulting from adjacent phosphate sites.  The model is also only capable of representing structures involving B-DNA and ssDNA, and the equal groove size in the model may also mask subtle effects related to major and minor grooving.

Ignoring sequence heterogeneity dramatically lowers the number of parameters needed for the coarse-grained model.  It also simplifies the analysis of the physical processes, naturally generating results for an "average strand".  This picture may be particularly advantageous when sequence effects obscure an important general trend.      Of course there are also many processes where sequence heterogeneity is critical, for example, preferred sites for bubble nucleation. Such  effects are not resolved by our model.     Nevertheless, for many applications in DNA nanotechnology, sequence dependent effects beyond complementarity are not that critical to design or functionality.    For example, in Ref.\ \onlinecite{Ouldridge_tweezers_2010} we show that the entropy cost of bringing an anti-fuel strand in close proximity to the tweezer complex slows down the displacement-mediated detachment of the first arm of the fuel strand.  Such predictions should be fairly robust and independent of sequence heterogeneity effects.    We also show how metastable hairpin formation in the anti-fuel strand can further affect the free-energy profile and the related kinetics of the displacement process.   Again, this general prediction should be fairly robust, but how it plays out for a particular set of tweezers will depend on how easily the anti-fuel strand sequence forms hairpins.   For example, if the metastable hairpin formation is undesirable, then our predictions could be supplemented by methods such as the nearest-neighbour model in order to design strands that minimize hairpin formation.

Similarly, in the current paper we make a series of predictions that should be relevant to experiment.  For example, we predict that a maximum in the magnitude of the enthalpy change of duplex formation, $\Delta H$, should occur as the temperature nears the polynucleotide melting temperature and fraying begins to reduce the number of base pairs in the bound state.  The exact  location and magnitude of the maximum will depend on sequence-dependent effects such as the exact melting temperature, whether the end bases form weaker AT or stronger CG bonds that promote or repress fraying, respectively, as well as the thermodynamics of single-stranded stacking.  But our prediction of a maximum in the absolute value of  $\Delta H$ should be fairly robust.

We also predict that the $T_m$ of a duplex containing a destabilizing motif should depend on the location of the motif in a temperature-dependent fashion. As the destabilizing motif is moved towards the centre of a duplex, the melting temperature should decrease before reaching a plateau. The distance from the end at which the plateau is observed will increase with $T_m$ and the destabilizing effect of the motif. Both of these effects result from sufficiently generic properties that we expect them to be resilient to the approximations of the model.

When compared to the nearest-neighbor model, our model tends to slightly underpredict the effect  of dangling ends, bulges, terminal mismatches and internal mismatches on the duplex melting temperature.  Again, it should be kept in mind that for real DNA the effect of each of these motifs will depend very much on the exact sequence, whereas our predictions are for an average over all possible sequence permutations.    Nevertheless,  in a system where multiple kinetic traps are relevant, extra care should be taken when interpreting the simulations because the relative stabilities of different states could be somewhat misrepresented.

We also make some predictions for the conformational statistics of dsDNA, suggesting that single strands behave much more like freely-jointed chains with excluded volume than like worm-like chains.
 A particular consequence of this difference is that freely-jointed chains typically undergo much larger local kinking than worm-like chains with an equivalent effective persistence length.


Finally, we have demonstrated that a nearest-neighbour two-state model of duplex formation can be extended to incorporate stacking and fraying. This extension suggests a way to reconcile the appealing simplicity of nearest-neighbour models with temperature variation of both single- and double-stranded states. To develop such a model, however, would require a much greater consensus in the properties of single-stranded stacking and fraying than currently exists. 

As it stands, we believe the model has the potential (both in terms of accuracy and computational efficiency) to open up a range of previously inaccessible general problems involving the interplay between single- and double-stranded DNA, including many aspects of  DNA nanotechnology. For example, we are investigating the operation of a  DNA walker,\cite{Bath2009} the force-induced melting of DNA,\cite{vanMameren2009} the assembly of a DNA tetrahedron\cite{Goodman2005} and binding of hairpins in the presence of a DNA catalyst.\cite{Bois2005}   The model may also be applied to biologically relevant processes such as the extrusion of cruciforms in supercoiled DNA containing inverted repeats.\cite{Sinden1994}
Future work will aim to incorporate sequence-dependent interaction strengths (we note that much of the sequence dependence should arise from stacking), major and minor grooving and an implicit model for electrostatics, as well as comparing to atomistic simulations to improve the description of fluctuations on the base pair level.

\bibliography{Bibliography_central.bib}

\appendix
\section{Model details and parameterization}
\label{appendix-model}
The current model is based on that introduced in Ref.\ \onlinecite{Ouldridge_tweezers_2010}, with some changes introduced to give duplexes more flexibility (having performed a wider range of structural tests, the stiffness was found to be overestimated in the old version). Truncated interactions have also been quadratically smoothed (making the potential continuous and differentiable, allowing simulation with methods like Langevin dynamics). Although this introduces further parameters, the thermodynamic and structural properties are largely unaffected by the details of smoothing.

The functional forms used in the interactions are given below:
\begin{itemize}
\item FENE spring (used to connect backbones):
\begin{equation}
V_{\rm fene}(r) = - \frac{k}{2} \ln \left( 1- \frac{(r-r_0)^2}{\Delta^2} \right).
\end{equation}
\item Morse potential (used for stacking and H-bonding):
\begin{equation}
V_{\rm Morse}(r, \epsilon, r_0,a) = \epsilon \big(1-\exp{(-(r-r_0)a)}\big)^2.
\end{equation}
\item Harmonic potential (used for cross-stacking):
\begin{equation}
V_{\rm harm}(r, \epsilon, r_0) =  \frac{\epsilon}{2} \left(r - r_0 \right)^2.
\end{equation}
\item Lennard - Jones potential (used for soft repulsion);
 \begin{equation} 
V_{\rm LJ}(r, \epsilon, \sigma) = 4\epsilon \left (\left({\sigma \over r} \right) ^{12} - \left( {\sigma \over r} \right) ^{6} \right) .
\end{equation}
\item Quadratic terms (used for modulation)
\begin{equation}
V_{\rm mod} (\theta, a, \theta_0) = 1 - a (\theta-\theta_0)^2
\end{equation}
\item Quadratic smoothing terms:
\begin{equation}
V_{\rm smooth} (x, b, x_c) = b(x_c - x)^2
\end{equation}
\end{itemize}

\begin{table*}[!]
\begin{center}
\begin{tabular}{c  c  c  c  c  c}
\hline
\hline
& & & &\\
{\bf Interaction} & {\bf Functional form} & \multicolumn{4}{c}{  \bf Parameters }\\\
\\ \hline 
& & & & \\
backbone spring & $V_{\rm fene}(r_{\rm backbone})$ & $k = 2$ & $\Delta= 0.25$ & $r_0 = 0.7525$ & \\ 
$V_{backbone}$ & & & & \\
\hline
& & & & \\
hydrogen bond & $f_1(r_{\rm bond})$ & $\epsilon = 1.077$ & $a = 8$ & $r_0 = 0.4$ & $r^{low} = 0.34$ \\
$V_{HB}$  & & & & $r_c=0.75$ & $r^{high} = 0.70$ \\
& $f_4(\theta_1)$ & $a = 1.50$ & $\theta_0 = 0$ & $\Delta \theta^{\star} = 0.70$ & \\
& $f_4(\theta_2)$ & $a= 1.50$ & $\theta_0 = 0$ & $\Delta \theta^{\star}= 0.70$ & \\
& $f_4(\theta_3)$ & $a= 1.50$ & $\theta_0 = 0$ & $\Delta \theta^{\star} = 0.70$ & \\
& $f_4(\theta_4)$ & $a = 0.46$ & $\theta_0 =  \pi$ & $\Delta \theta^{\star} = 0.70$ & \\ 
& $f_4(\theta_7)$ & $a= 4.00$ & $\theta_0 = \pi/2$ & $\Delta \theta^{\star} = 0.45$ & \\
& $f_4(\theta_8)$ & $a = 4.00$ & $\theta_0 =  \pi/2$ & $\Delta \theta^{\star} = 0.45$ & \\ \hline
& & & & \\
stacking & $f_1(r_{\rm stack})$ & $\epsilon = 1.2145$ & $a = 6$ & $r_0 = 0.4$ &  $r^{low} = 0.32$\\
$ V_{stack}$ & &$ + 2.6568\,{kT}$  & & $r_c=0.9$ &  $r^{high} = 0.75$\\
& $f_4(\theta_4)$ & $a = 1.30$ & $\theta_0 = 0$ & $\Delta \theta^{\star}= 0.8$ & \\
& $f_4(\theta_5)$ & $a = 0.90$ & $\theta_0 = 0$ & $\Delta \theta^{\star} = 0.95$ & \\
& $f_4(\theta_6)$ & $a = 0.90$ & $\theta_0 = 0$ & $\Delta \theta^{\star}= 0.95$ & \\ 
& $f_5(\cos(\phi_1))$ & $a=2.00$ &   $x^{\star}= -0.65$ & \\ 
& $f_5(\cos(\phi_2))$ & $a=2.00$ &  $x^{\star}= -0.65$ & \\ \hline
& & & &  \\
cross-stacking & $f_2(r_{\rm cstack})$ & $\epsilon = 47.5$ & $r_0 = 0.575$ & $r_c=0.675$ & $r^{low} = 0.495$\\
$V_{c\_stack}$ & & & & & $r^{high} = 0.655$\\
& $f_4(\theta_1)$ & $a = 2.25$ & $\theta_0 = 2.35$ & $\Delta \theta^{\star} =0.58$ & \\
& $f_4(\theta_2)$ & $a = 1.70$ & $\theta_0 = 1.00$ & $\Delta \theta^{\star} =0.68$ & \\
& $f_4(\theta_3)$ & $a = 1.70$ & $\theta_0 = 1.00$ & $\Delta \theta^{\star} =0.68$& \\ 
& $f_4(\theta_4)+f_4( \pi - \theta_4)$ & $a = 1.50$ & $\theta_0 = 0$ & $\Delta \theta^{\star} =0.65$ & \\
& $f_4(\theta_7) + f_4(\pi - \theta_7)$ & $a=1.70$ & $\theta_0 = 0.875$ & $\Delta \theta^{\star} =0.68$ & \\
& $f_4(\theta_8) + f_4(\pi - \theta_8)$ & $a = 1.70$ & $\theta_0 = 0.875$ & $\Delta \theta^{\star} =0.68$& \\\hline
& & & & \\
excluded volume & $f_3(r_{\rm ex1})+f_3(r_{\rm ex2})$ & $\epsilon = 2.00$ & $\sigma_1 = 0.70$ & $r_1^{\star} =0.675$  & \\
$V_{exc}$ & $+f_3(r_{\rm ex3})+f_3(r_{\rm ex4})$ & & $\sigma_2 = 0.33$& $ r_2^{\star} =0.32$ &  \\ 
& & & $\sigma_3 = 0.515$ &  $r_3^{\star} =0.50$ &\\
& & & $ \sigma_4 = 0.515$ & $ r_4^{\star} =0.50$ & \\
\hline
\hline
\end{tabular}
\caption{Parameter values in the model.
All lengths are defined with respect to a reduced lengthscale (1 unit = 8.518\AA), all angles are given in radians and all energies are defined with respect to a reduced temperature ($kT = 0.1$ corresponding to 300\,K).
The variables in the potential are defined in Fig.\ \ref{diagram}}
\label{parameters}
\end{center}
\end{table*}

These functional forms are combined to give the following smooth and differentiable functions:
\begin{widetext}
\begin{itemize}
\item The radial part of the stacking and hydrogen-bonding potentials:
\begin{equation}
f_1(r) = \begin{cases}
	V_{\rm Morse}(r, \epsilon, r_0, a) - V_{\rm Morse}(r_{c}, \epsilon, r_0, a)) & \text{if $ r^{low} < r < r^{high} $},\\
	\epsilon V_{\rm smooth} (r, b^{low}, r_c^{low}) & \text{if $r_c^{low} < r < r^{low}$},\\
	\epsilon V_{\rm smooth} (r, b^{high}, r_c^{high}) & \text{if $r^{high} < r < r_c^{high}$},\\
	0 & \text{otherwise}.
	\end{cases} 
\end{equation}
\item The radial part of the cross-stacking potential:
\begin{equation}
f_2(r) = \begin{cases}
	V_{\rm harm}(r, \epsilon, r_0) - V_{\rm harm}(r_{c}, \epsilon, r_0) & \text{if $ r^{low} < r < r^{high} $},\\
	\epsilon V_{\rm smooth} (r, b^{low}, r_c^{low}) & \text{if $r_c^{low} < r < r^{low}$},\\
	\epsilon V_{\rm smooth} (r, b^{high}, r_c^{high}) & \text{if $r^{high} < r < r_c^{high}$},\\
	0 & \text{otherwise}.
	\end{cases} 
\end{equation}
\item The radial part of the excluded volume potential:
\begin{equation}
f_3(r) = \begin{cases}
	V_{\rm LJ}(r, \epsilon, \sigma) & \text{if $r < r^{\star} $},\\
	\epsilon V_{\rm smooth} (r, b, r_c) & \text{if $r^{\star} < r < r_c$},\\
	0 & \text{otherwise}.
	\end{cases} 
\end{equation}
\item The angular modulation factor used in stacking, hydrogen bonding and cross-stacking:
\begin{equation}
f_4(\theta) = \begin{cases}
	V_{\rm mod}(\theta, a, \theta_0)  & \text{if $ \theta_0 - \Delta \theta^{ \star} < \theta < \theta_0 + \Delta \theta^{\star} $},\\
	V_{\rm smooth} (\theta, b, \theta_0 - \Delta \theta_c) & \text{if $\theta_0 - \Delta \theta_c < \theta < \theta_0-\Delta \theta^{\star}$},\\
	V_{\rm smooth} (\theta, b, \theta_0 + \Delta \theta_c) & \text{if $\theta_0 + \Delta \theta^{\star} < \theta < \theta_0+\Delta \theta_c$},\\
	0 & \text{otherwise}.
	\end{cases} 
\end{equation}
\item Another modulating term which is used to impose right handedness (effectively a one-sided modulation):
\begin{equation}
f_5(\phi) = \begin{cases}
	1 & \text{if $x > 0$},\\
	V_{\rm mod} (x, a, 0) & \text{if $x^{\star} < x < 0$},\\
	V_{\rm smooth} (x, b, x_c) & \text{if $x_c< x <x^{\star}$},\\
	0 & \text{otherwise}.
	\end{cases} 
\end{equation}
\end{itemize}

\end{widetext}

\begin{figure*}
\begin{center}
\resizebox{120mm}{!}{\includegraphics{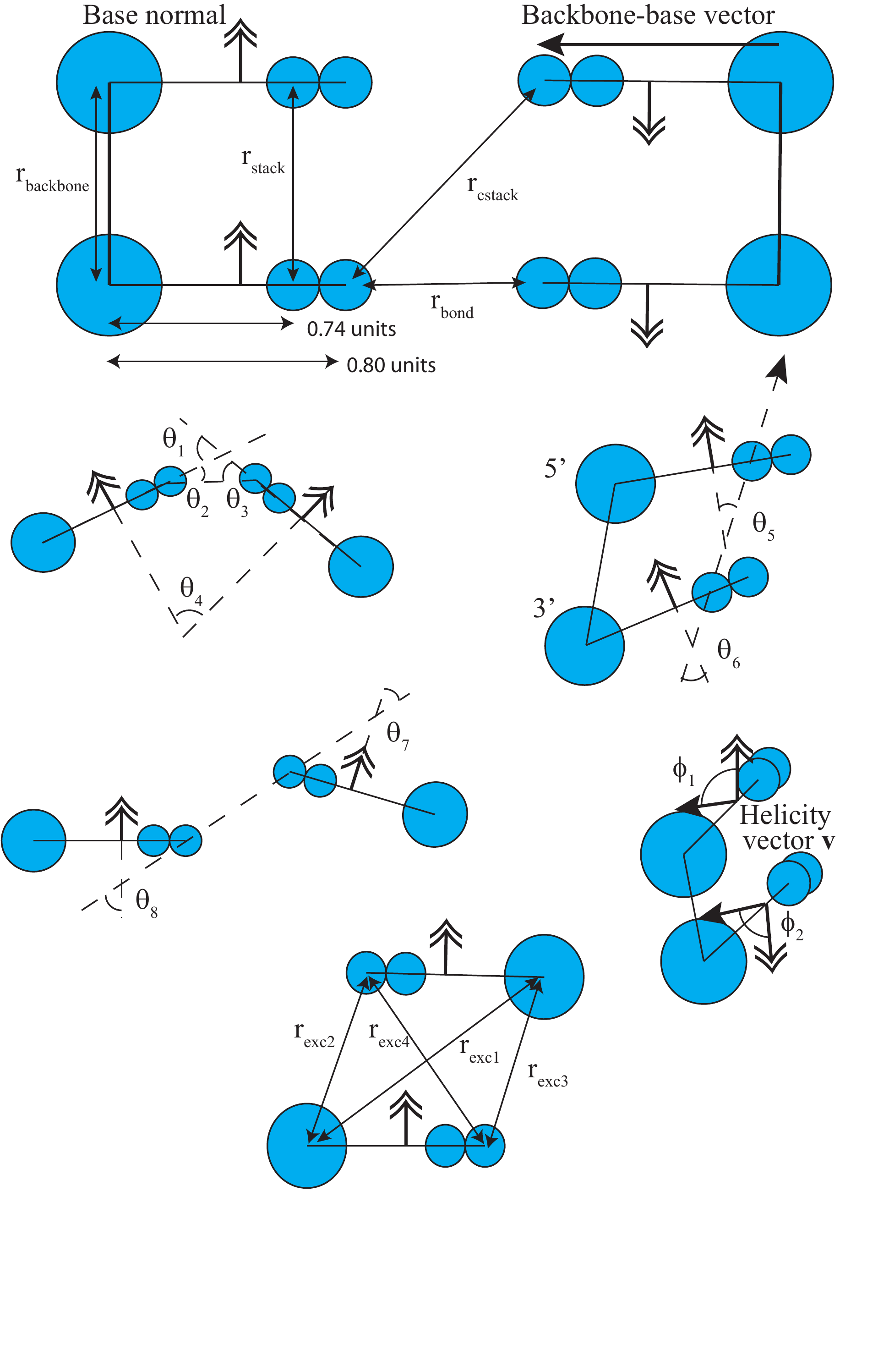}}
\caption{Illustration of variables used in the potential of the DNA model.}
\label{diagram}
\end{center}
\end{figure*}

The potentials and parameters used to describe each interaction are listed in the Table.\,\ref{parameters}.  When more than one function is listed for an interaction, the total interaction is a product of all the terms. Given the parameters of the main part of the interaction (for example, $\epsilon$, $r_0$, $a$ and $r_c$ for the $V_{\rm Morse}$ part of $f_1(r)$), the parameters of the smoothed cutoff regions are uniquely determined by  ensuring continuity and differentiability at the boundaries ($r^{low}$ and $r^{high}$ for $f_1(r)$).
The nucleotide geometry and definition of the angles and vectors used in the potential are shown in Fig.\ \ref{diagram}.

The potential of the system is given by:
\begin{equation}
\begin{array}{cc}
V & =  \dSum_{\rm nn} \big(V_{backbone}  + V_{stack}+V^{\prime}_{exc}\big) \\ 
& +\dSum_{\rm other \,pairs} \big(V_{HB} + V_{c\_stack}+V_{exc}\big),
\end{array}
\end{equation}
where the sum over nn runs over consecutive bases within strands, and $V^\prime_{exc}$ is equal to $V_{exc}$ except that it does not include an $f_4(r_{\rm ex1})$ term. Note the directional dependence in the stacking interaction: the angles are defined between normal vectors of bases and a vector joining bases in the 3' to 5' direction. Only complementary base pairs posses non-zero hydrogen-bond energies. 

To ensure right-handed helices, the modulation of stacking interactions is somewhat subtle, involving chiral terms.
Consider two consecutive bases in a strand, $i$ and $j$, with $i \rightarrow j$ corresponding to the $3' \rightarrow 5'$ direction. The angles $\theta_5$ and $\theta_6$ are defined as the angles between the normals of $i$ and $j$ and $\bf{r}^{ij}_{\rm stack}$ (with $\bf{r}^{ij}_{\rm stack}$ being defined as the vector from the stacking site of $i$ to that of $j$). Thus, stacked bases have normals pointing in the $3' \rightarrow 5'$ direction, allowing the definition of a local axis.
The angles $\phi_1$ and $\phi_2$, defined in terms of this axis, provide helicity. For each base we define a normalized vector ${\text{\bf \^v}}^{\alpha}_{\rm helicity} = {\text{\bf \^r}}^{ij} \times{ \text{\bf  \^r}}^{\alpha}_{\rm back-base}$, where $\alpha = i,j$ and ${\text{\bf \^ r}}^{\alpha}_{\rm back-base}$ is the normalized backbone site to stack site vector of base $\alpha$. $\phi_1$ and $\phi_2$ are the angles between ${\text{\bf \^v}}^{\alpha}$ and the base normals: for a right handed helix, these are $<\pi/2$ and the stacking interaction is modulated to disfavour greater angles.

\section{Statistical model of stacking}
\label{appendix-stacking}
It is instructive to characterize the thermodynamics of the model using a simpler, statistical model, as it highlights the causes of certain behaviour. We model the stacking transition using a statistical description based on that of Poland and Scheraga.\cite{Poland1970} In this model, a given pair of neighbours can be either stacked or unstacked, and the list of stacked pairs specifies the system configuration. 

If each stacking pair were independent, the contribution to the partition function from a configuration (its relative probability of occurring) would be given by:
\begin{equation}
Z_{\rm config} =  z_0 u^{N_i}  v^{N_j}
\label{independent}
\end{equation}
where $u$ and $v$ represent the contributions to the partition function (``statistical weight") of a stacked and an unstacked pair respectively, $N_i$ and $N_j$ are the number of stacked an unstacked pairs and $z_0$ denotes the trivial contribution from translation and orientation of the whole strand. As discussed in Section\,\ref{stacking}, the excluded volume of nucleotides means that pairs of neighbours are not independent. To deal with this, we introduce two new parameters. The statistical weight of a continuous section of $n$ stacked pairs is now given by:
\begin{equation}
u({n}) = \sigma u^{n} w^{x},
\end{equation}
with $x$ being equal to the number of bases in the run of stacked pairs that lie at the end of the strand. $n$ unstacked pairs contribute the same statistical weight as before:
\begin{equation}
v({n}) =v^{n}.
\end{equation}
If $\sigma$ and $w$ are unity, each neighbour pair is independent, and we return to Eqn.\ \ref{independent}. $\sigma$ takes the role of a cooperativity parameter: for $0 < \sigma <1$, stacking is cooperative, in that configurations with multiple separate regions of stacking are disfavoured, and for $\sigma >1$ stacking is anticooperative. $w$ accounts for end effects: for $0 < w <1$, end bases are less likely to stack, and for $w>1$ the opposite is true. 

Using these definitions, the total partition function for a strand of length $l$ becomes:
\begin{equation}
Z_{l} =  \sum _{\{n_i, m_j; l\}} z_0 w^x \prod_i \sigma u^{n_i} \prod_j v^{m_j}.
\label{stack partition 2}
\end{equation}
Here, $\{n_i, r_j; l\}$ specifies a configuration, $n_i$ being the number of stacked pairs in the $i^{\rm th}$ contiguous sequence of stacked neighbours, $m_j$ being the number of unstacked pairs in the $m^{\rm th}$ sequence of unstacked bases and $x= \sum_i x_i$ is the total number of bases at the end of the strand involved in stacking. 

Defining $t = u / v$, $n= \sum_i n_i$ and letting $p$ be the total number of stacked regions, we obtain:
\begin{equation}
Z_{l} =  Z_l^u\sum _{\{n_i, m_j; l\}} w^x \sigma^p t^n.
\label{stack partition 3}
\end{equation}
with $Z_l^{u} = z_0 v^{l-1}$ being the partition function of a completely unstacked strand.
To compare directly with simulations, we require the ratio of the probability of observing $r$ stacked pairs to the probability of observing a completely unstacked strand:
\begin{equation}
\frac{Z_{l} (r)}{Z_{l}^u} =  \sum _{\{n = r; l \}} w^x \sigma^p t^n  =t^r \sum_{x=0}^2 w^x  \sum_p \sigma^p \Omega_{\{x,r,p;l\}},
\label{stack partition 4}
\end{equation}
with $\Omega_{\{x,r,p;l\}}$ defined as the number of distinct configurations of length $l$ with $r$ stacked pairs, of which $x$ are at the end of the strand, divided between $p$ contiguous regions of stacking. The advantage of this representation is that finding $\Omega_{\{x,r,p;l\}}$ is simply a matter of combinatorics.
It can be shown that: 
\begin{equation}
\Omega_{\{x,r,p;l\}} = \frac{(1 + \delta^x_1)(r-1)!(l-r-2)!}{(r-p)! (p-1)!(l-r-2-p+x)! (p-x)!} ,
\end{equation}
for all possible values of $x$, $r$ and $p$ for a strand of length $l$, 
with the exception that  $\Omega_{\{0,0,0;l\}} = \Omega_{\{2,l-1,1;l\}} = 1$.

We assume that the temperature dependence of stacking is manifested in the parameter $t$, which is defined as $ t = \exp(- \Delta h^{st} / {R T} + \Delta s^{st} / {R})$, with $\Delta h^{st}$  and $\Delta s^{st}$ representing the (assumed constant) enthalpy and entropy changes associated with stack formation. As $w$ and $\sigma$ arise from excluded volume effects, they are assumed to be entropic and hence temperature independent. We fitted this 4-parameter model to data obtained in simulations, the results are shown in Section\,\ref{stacking}.

\section{Statistical model for duplex formation}
\label{appendix-duplex}
Eqn.\ \ref{2smb} assumes a constant entropy and enthalpy difference between bound and unbound states. It is well known, however, that $\Delta S_l$ and $\Delta H_l$ should both become more negative with temperature, as the unbound strands become increasingly disordered due to unstacking.\cite{Vesnaver1991, Holbrook1999, Mikulecky2006,  Jelesarov1999} Using the formalism of Appendix \ref{appendix-stacking}, we can factor out this effect:
\begin{equation}
\frac{[A_l B_l]}{[A_l][B_l]}  = v \frac{Z_{ll}}{Z_l^2}= \frac{ \exp\Big(-\beta \big(\Delta H'_l - T \Delta S'_l\big)\Big) (Z_l^{u})^2}{Z_l^2},
\label{3sm}
\end{equation}
where in this case $\Delta H^\prime_l$ and $\Delta S^\prime_l$ are the enthalpy and entropy difference between the duplex and unstacked single-stranded macrostates. 

Although fitting to Eqn.\ \ref{3sm} with constant $\Delta H^\prime_l$ and $\Delta S^\prime_l$ was more successful than assuming constant $\Delta H_l$ and $\Delta S_l$, it overcorrected for the variations in $\Delta S_l$ and $\Delta H_l$ with temperature. The failure resulted from neglecting the changes in the bound state with temperature, which were dominated by two effects:
\begin{itemize}
\item As temperature increases, increased fraying leads to smaller entropy and enthalpy differences between typical bound states and completely unstacked single strands, as bound states become more disordered.
\item Frayed ends themselves undergo a stacking transition, once more resulting in the entropy and enthalpy of bound states relative to unstacked strands becoming less negative with temperature.
\end{itemize}

\begin{figure}
\begin{center}
\resizebox{80mm}{!}{\includegraphics[angle=-90]{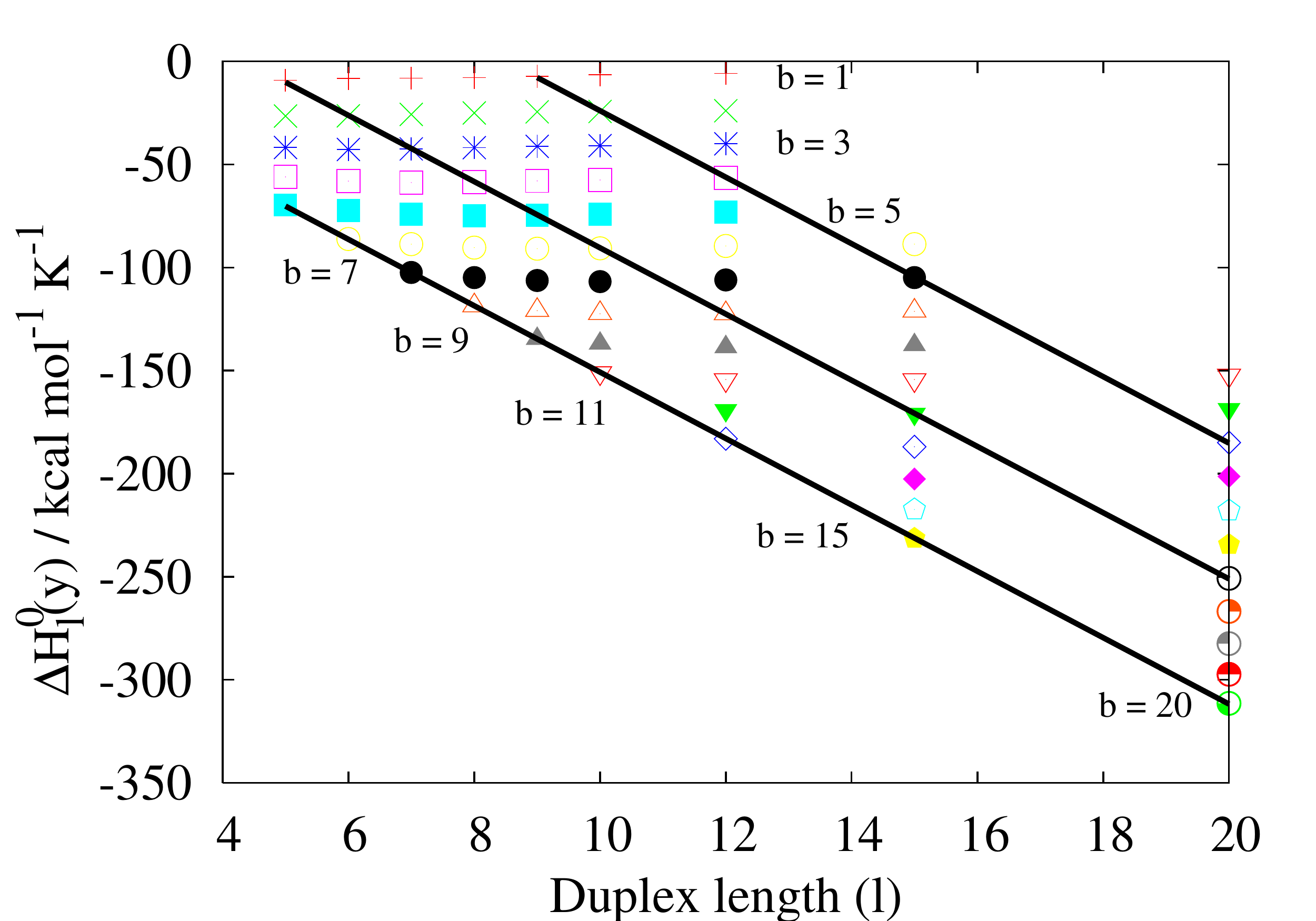}}
\end{center}
\caption{\footnotesize $\Delta H^0_l(y)$ against duplex length $l$. Points are styled according to the total number of bonds formed, $b=l-y$. The solid lines are linear fits for the dependence of $\Delta H^0_l(y)$ on $l$ for fixed $y$.}
\label{msm-graph}
\end{figure}

To incorporate these effects within a statistical model, we separately consider the entropy and enthalpy differences between unstacked single strands and macrostates with $y$ out of $l$ possible base pairs 
formed. We then approximately adjust for stacking of the frayed ends by treating the $2(l-y)$ unpaired bases as undergoing stacking with the same $\Delta h^{st}$ and $\Delta s^{st}$ given in Section \ref{stacking}. Cooperativity and end effects are ignored as it would be difficult to include them consistently when stacking is initiated adjacent to a duplex region.

We thus define $Z_{ll}(y)$ as:
\begin{equation}
Z_{ll} = \sum_y Z_{ll}(y), 
\label{Z_ll(y)}
\end{equation}
and $Z^u_{ll}(y)$ as the contribution to $Z_{ll}(y)$ in which none of the unpaired bases are stacked. $Z^u_{ll}(y)$  is approximated by:
\begin{equation}
Z_{ll}(y) =  Z^u_{ll}(y) \Big(1+ \exp \big(- \beta (\Delta h^{st} - T \Delta s^{st}) \big) \Big)^{2(l-y)},
\label{Zll_Zllu}
\end{equation}

Our hypothesis is that the enthalpy and entropy differences between unstacked single strands and the states contributing to $Z^u_{ll}(y)$ should be approximately constant for given $l$ and $y$, as the temperature variation due to breaking stacks and fraying has been factored out. The values of ${Z_{ll}(y)}/{Z_l^2}$ were extracted from the fraying data, and ${Z^u_{ll}(y)}/{(Z^u_l)^2}$ inferred using Eqns.\ \ref{stack partition 4} and \ref{Zll_Zllu}. Fitting to
\begin{equation}
v \frac{Z^u_{ll}(y)}{(Z^u_l)^2}= \exp\Big(-\beta \big(\Delta H^0_l(y) - T \Delta S^0_l(y)\big)\Big)
\label{msm}
\end{equation}
with constant $\Delta H^0_l(y)$ and $\Delta S^0_l(y)$ (which represent the enthalpy and entropy differences between unstacked single strands and the states contributing to $Z^u_{ll}(y)$) was very successful. 

Furthermore, as shown in Fig.\ \ref{msm-graph}, $\Delta H^0_l(y)$ (and $\Delta S^0_l(y)$, which is not shown) are to an excellent approximation linear in $l$ for fixed $y$. Thus, having factored out sources of variation with temperature in the initial and final states, we arrive at a statement similar to the initial hypothesis of the nearest-neighbour model: adding an extra bp to a helix (i.e., increasing the length of the strands by one base, and forming one extra base pair, so that the number of unpaired bases is constant) contributes a constant enthalpy and entropy change relative to unstructured single strands. 

This finding suggests an extension of the nearest-neighbour model to non-two-state behaviour to incorporate fraying and stacking, and thus predict the values of $\Delta S(T)$ and $\Delta H(T)$ for oligonucleotides. To achieve this description, fraying and stacking transitions must be sufficiently well characterized, and the assumption that helix stability is predominantly due to nearest-neighbour effects must hold, as it does in our model. It should be noted that at low temperatures, certain oligomers may also have significant contributions to the single-stranded state from hairpins, which are not incorporated into this model. 

We are finally in a position to characterize the hybridization transition with completely temperature independent parameters. Combining Eqns.\ \ref{2smb},\,\ref{stack partition 4},\,\ref{Z_ll(y)},\,\ref{Zll_Zllu} and\,\ref{msm}, we find:
\begin{widetext}
\begin{equation}
K_{eq} = \exp\Big(-\beta \big(\Delta H_l - T \Delta S_l\big)\Big) = v \frac{Z_{ll}}{Z_l^2} = \frac{\sum_y \exp\Big(-\beta \big(\Delta H^0_l(y) - T \Delta S^0_l(y)\big)\Big) \Big(1+ \exp \big(- \beta (\Delta h^{st} - T \Delta s^{st}) \big) \Big)^{2(l-y)}}{\sum_r \exp \big(- \beta (\Delta h^{st} - T \Delta s^{st}) \big) ^r  \sum_{x}^2 w^x \sum_p \sigma^p \Omega_{\{x,r,p;l\}}}
\end{equation}
\begin{equation}
\Delta H_{l} = -\frac{\rm d}{\rm d \beta} \ln K_{eq} = \frac{\sum_y \left(\Delta H^0_l(y) + 2(l-y) \Delta h^{st} \frac{\exp (- \beta (\Delta h^{st} - T \Delta s^{st})}{1+ \exp (- \beta (\Delta h^{st} - T \Delta s^{st}))}\right)Z_{ll}(y)}{Z_{ll}} - 2 \frac{\sum_r \left( r \Delta h _{st} Z_l(r)\right)}{Z_l}.
\label{dH(T)_eqn}
\end{equation}
\end{widetext}
Eqn.\ \ref{dH(T)_eqn} is used in Section\,\ref{dH(T)} to produce the fit of $\Delta H_{15}$ to simulations. The first term gives the enthalpy of duplexes with respect to unstacked single strands and the second term the enthalpy of two single strands with respect to their unstacked state. As can be seen, the agreement is good over  a wide range of temperatures. Had hairpins been possible in the simulations of Section\,\ref{dH(T)}, they may have distorted the enthalpy at temperatures far below $T_m$. Hairpins were excluded to make the interpretation of results clearer, and their presence would have made the single-stranded state's enthalpy more negative. This change would have led to a smaller transition enthalpy between single strands and duplexes at these temperatures.

\end{document}